\documentclass[10pt-default]{article}
\usepackage{amsmath}
\usepackage{amssymb}
\usepackage{amsfonts}
\usepackage[dvips]{graphicx}
\usepackage{amsfonts}

\input{tcilatex}

\begin{document}

\bigskip\thinspace{\huge Veneziano amplitudes, }

\ {\huge spin chains and }

{\huge Abelian reduction of QCD }

$\ \ \ \ \ \ \ \ \ \ \ \ \ \ \ \ \ \ \ \ \ \ \ \ \ \ \ \ \ \ \ \ \ $

\ \ \ \ \ \ \ \ \ \ \ \ \ \ \ Arkady Kholodenko\footnote{%
\bigskip
\par
Tel.1-864-656-5016;
\par
fax 1-864-656-6613
\par
E-mail address: string@clemson.edu}

\textit{375 H.L.Hunter Laboratories, Clemson University,}

\ \ \ \ \ \ \ \ \ \ \ \ \ \ \textit{Clemson, \ SC} 29634-0973, U.S.A.

\textbf{Abstract}

Although QCD can be treated perturbatively in the high energy limit, lower
energies require uses of nonperturbative methods such as ADS/CFT \ and/or
Abelian reduction. These methods are not equivalent. While the first is
restricted to supersymmetric Yang-Mills\ model with number of colors going
to infinity, the second is not restricted by requirements of supersymmetry
and is designed to work in physically realistic limit of finite number of
colors. In this paper we provide arguments in favor of the Abelian reduction
methods. This is achieved by further developing results of our recent works
reanalyzing Veneziano and Veneziano-like amplitudes and the models
associated with these amplitudes. It is shown, that the obtained \ new
partition function for these amplitudes can be mapped \ exactly into that
for the Polychronakos-Frahm (P-F) spin chain model recoverable from the
Richardon-Gaudin (R-G) XXX \ spin chain model originally designed for
treatments of the BCS-type superconductivity. Because of this, it is
demonstrated that the obtained mapping is compatible with the method of
Abelian reduction. The R-G model is recovered \ from the asymptotic
(WKB-type) solutions of \ the rational Knizhnik-Zamolodchikov (K-Z)
equation. Linear independence of \ these solutions is controlled by
determinants whose explicit form (up to a constant) coincides with Veneziano
(or Veneziano-like) amplitudes. In the simplest case, the determinantal
conditions coincide with those discovered by Kummer in 19-th century.
Kummer's results admit physical interpretation by relating determinantal
formula(s) to Veneziano-like amplitudes. Furthermore, these amplitudes \ can
be interpreted as Poisson-Dirichlet distributions playing central role in
the stochastic theory of random coagulation-fragmentation processes. Such an
interpretation is complementary to that known for the Lund model widely used
for description of coagulation-fragmentation processes in QCD.



\section{Introduction}

Since times when quantum mechanics (QM) was born (in 1925-1926) two
seemingly opposite approaches for description of atomic and subatomic
physics were proposed respectively by Heisenberg and Schr\"{o}dinger.
Heisenberg's approach is aimed at providing an affirmative answer to the
following question: Is combinatorics of spectra (of obsevables) \ provides
sufficient \ information about microscopic system so that dynamics of such a
system can be described in terms of known macroscopic concepts?
Schrodinger's approach is exactly opposite and is aimed at providing an
affirmative answer to the question: Using some plausible mathematical
arguments is it possible to find equations which under some prescribed
restrictions will reproduce the spectra of observables? Although it is
widely believed that both approaches are equivalent, already Dirac in his
lectures on quantum field theory, Ref.[1], noticed (without much
elaboration) that Schrodinger's description of QM contains a lot of "dead
wood" which can be safely disposed. According to Dirac \ "Heisenberg's
picture of QM is good because Heisenberg's equations of motion make sense".

To our knowledge, Dirac's comments were completely ignored, perhaps, because
he had not provided enough evidence making Heisenberg's description of QM
superior to that of Schrodinger's. In recent papers, Ref.[2,3], we found
examples supporting Dirac's claims. From the point of view of combinatorics,
there is not much difference in description of QM, quantum field theory or
string theory as demonstrated in Ref.[4]. Therefore, in this paper we adopt
Heisenberg's point of view at QCD and string theory using results of our
recent works aimed at re analyzing the existing treatments connecting
Veneziano (and Veneziano-like) amplitudes with the respective
string-theoretic models. As result, we found \ new tachyon-free models
reproducing Veneziano (and Veneziano-like) amplitudes. In this work results
of our \ papers, Ref.s.[5-7], to be called respectively as Part I, Part II
and Part III are developed further. This has become possible in view of the
work by Reshetikhin and Varchenko, Ref.[8], \ and by Varchenko summarized in
Varchenko's \ MIT lecture notes, Ref.[9]. They enabled us to relate
Veneziano (and Veneziano-like) amplitudes (e.g.those describing $\pi \pi $
scattering) to Knizhnik-Zamolodchikov (K-Z) equations and, hence, to WZNW
models. This is achieved by employing known connections between WZNW models
and spin chains. In the present case, between the K-Z equations and the
XXX-type Richardson-Gaudin (R-G) spin chains. To \ keep things in a
perspective, we would like to provide some rationale behind the above
mentioned connection with spin chains developed in this paper.

As is well known, all information in high energy physics is obtainable
through proper interpretation of scattering data. It is believed that for
sufficiently high energies such data are well described by the
phenomenological Regge theory and can be conveniently summarized with help
of Chew-Frauthchi (C-F) plots relating masses to spins (angular momenta),
e.g. see book by Collins, Ref.[10]. Veneziano amplitudes are by design
Regge-behaving. Both Regge theory and Veneziano amplitudes emerged before
major developments in QCD took place in 70ies. Once these developments took
place, naturally, it was of interest to recover the Regge theory from QCD.
Even though there are many ways of doing so, to our knowledge, the problem
is still not solved completely. This is so for the following reasons.

Although the amount of data obtained by perturbative treatments of QCD is
quite impressive, e.g. read Ref.[11], these results are not helpful for
establishing the Regge-type behavior of QCD. \ Since such a behavior can be
easily established with help of variety of string models, the task lies is
connecting these models with QCD. For the sake of space, we do not discuss
broader reasons (e.g. quantum gravity) for development of string theory. \
Recently, another pathways towards establishing Regge-type behavior were
explored. For instance, in Ref.[12], in the large N limit (to be defined in
the main text) by ignoring masses of quarks the spin zero glueball mass
spectrum was obtained analytically for 2+1 dimensional pure Yang-Mills theory%
\footnote{%
These results were subsequently extended to 3+1 dimensions in Ref.[13].}
which in the high energy limit fits perfectly Regge theory. \ Although in
Ref.[12] string models were \textsl{not} used, the obtained results are in
excellent agreement with large N lattice calculations. Given this, they \
still suffer from several drawbacks. First, the large N limit, even though
mathematically convenient, physically is questionable as it will be
explained below, in the main text. Second, more physically interesting are
the spectra of mesons and baryons. These spectra are traditionally obtained
with help of string-theoretic models. Since calculations \ involving these
models are by design made Regge-behaving, the task was (still is) to connect
these string models with QCD.\ It is widely believed \ that such a
connection is achievable via ADS/CFT correspondence [14] between strings
living in anti-de Sitter space and $\mathcal{N}=4,$ N$\rightarrow \infty $
supersymmetric Yang-Mills (Y-M) model in $d=3+1$ dimensions. By design, such
supersymmetric Y-M model does not contain quark masses\footnote{%
The achievents of this method are summarized in excellent recent reviews by
Benna and Klebanov, Ref.[15], and Brodsky, Ref.[16]. \ Our work, Ref.[17],
provides an introduction to the ideas and methods of ADS/CFT.}. Again, such
correspondence becomes physically meaningful only if the number of colors N
could be made finite and the supersymmetry could be broken. \ Unfortunately,
these requirements are in apparent contradiction with the way the ADS/CFT
correspondence was established in the first place thus making such a task
very difficult to accomplish. \ 

In 1981 't Hooft suggested in Ref.[18] to reduce the non Abelian QCD to
Abelian Ginzburg-Landau (G-L) type theory. The rationale for such an \textsl{%
Abelian} \textsl{reduction} can be traced back to the work of Nambu,
Ref.[19]. In his work Nambu superimposed G-L theory with the theory of Dirac
monopoles to demonstrate quark confinement for mesons. Incidentally,
Veneziano amplitudes are suited the most for describing meson resonances,
e.g. see Ref.[10]. These are made of just two quarks: quark and antiquark.
Thus, if the existence of Abelian reduction would be considered as proven,
this then would be equivalent to the\ proof of quark confinement. Recent
numerical studies have provided convincing evidence supporting the idea of
quark confinement through monopole condensation, e.g. see Refs.[20,21].
Since publication of 't Hooft's paper many theoretical advancements were
made, most notably by Cho, Ref.s[22,23], and Kondo, Ref.s [24-26], whose
work was motivated \ by that by Faddeev and his group. Results of this group
are summarized in\ the recent review by Faddeev, Ref.[27]. From this
reference it follows that most of efforts to date were spent on description
of the massless version of QCD (just like in the case of ADS/CFT
correspondence, but without invoking supersymmetry or requiring N$%
\rightarrow \infty $). Excitation spectrum of solitonic knotted-like
structures (admitting interpretation in terms of closed strings) provides
the spectrum of glueball masses as demonstrated in Ref.[25]\footnote{%
It should be noted that the problem of existence of a mass gap in QCD was
included into seven the most outstanding millenium prize mathematical
problems of the 21st century by the Clay Mathematical Institute, e.g. see
http://www.claymath.org/millennium/}. In a recent paper, Ref.[28], Auckly,
Kapitanski and Speight demonstrated how Skyrme model can be obtained from
Faddeev model. Since Skyrme model was used for a long time for description
of the baryon spectra, e.g. read Ref.[29], and since already Nambu
recognized usefulness of the Abelian reduction for description of meson
spectra, it follows that the Abelian reduction method is capable of
providing sufficient information about QCD in the strong coupling regime.
Furthermore, in another paper, Ref.[30], Kapitanski and Auckley obtained
result of major importance for this paper. They demonstrated that it is
always possible to find such Chern-Simons (C-S) functional which upon
minimization will produce the same results as those obtained by minimization
of either Skyrme or Faddeev model. Asorey, Falceto and Sierra in Ref.[31]
demonstrated how C-S model is related to microscopic BCS \ model of
superconductivity. Since the Abelian reduction of QCD produces G-L-type
model (or collection of G-L models\footnote{%
E.g. read Faddeev's paper, Ref.[27].}), and since the underlying microscopic
model is of BCS-type whose elementary excitations are described by the
Richardson-Gaudin spin chain model, Ref.[31], the task of this work lies in
demonstrating that the combinatorics of scattering processes associated with
Veneziano (and/or Veneziano-like) amplitudes leads to the R-G spin chain
model of BCS superconductivity. The demonstrated in this work spin chain
connection made in the spirit of Heisenberg's work on quantum mechanics,
Ref.[2], favors the Abelian reduction method over ADS/CFT. \ Unlike other
Schr\"{o}dinger-style papers discussed above-all extracting the Abelian
Ginzburg-Landau-type\ model from the non Abelian QCD- the results of this
work use only combinatorics of scattering data as an input to arrive at the
same conclusions.

It should be noted that connections between either QCD and spin chains or
between strings\footnote{%
Within the context of ADS/CFT.} and spin chains were already discussed in
literature for quite some time. Recent paper by Dorey, Ref.[32], contains
may references \ listing these earlier results. Subsequently, they had been
replaced by those whose methods are based on ADS/CFT correspondence. From
the point of view of this correspondence, connections between strings and
QCD also can be made through spin chains as it is demonstrated in the
seminal paper by Gubser, Klebanov and Polyakov, Ref.[33]. Their ideas were
developed in great detail in the paper by Minahan and Zarembo, Ref.[34]. The
spectrum of anomalous dimensions of operators in the $\mathcal{N}=4,$ N$%
\rightarrow \infty $ supersymmetric Yang-Mills (Y-M) model (described in
terms of excitation spectrum of the spin chain model) is related to the
string spectrum describing hadron masses. To connect these facts with
developments in this paper we mention papers by Kruczenski, Ref.[35], and
Cotrone \textit{et al}, Ref.[36]. In both papers spin 1/2 XXX Heisenberg
chain was used for description of excitation spectrum. Furthermore, in the
paper by Cotrone\textit{\ et al }explicit connection with the hadron mass
spectrum was made. Both papers invoke ADS/CFT correspondence. In this work,
we reobtain these spin chain results using combinatorial arguments following
Heisenberg's philosophy\footnote{%
A review of Heisenberg's \ arguments resulted in birth of moden quantum
mechanics can be found in Ref.[2].}.

In Section 2 we reobtain Veneziano partition function (derived much more
rigorously in Part II). In Section 3 we demonstrate that \ this partition
function \ coincides with that for the Polychronakos-Frahm (P-F) spin chain
model. Although such a model was \ studied extensively in literature, in
Ref.[37] we discuss a variety of new pathways establishing links between the
P-F spin chain and many known string-theoretic models, including the most
recent ones. For the sake of space, we do not reproduce these results in
this work. Instead, in Section 4 we use results of Reshetikhin and
Varchenko, Ref.[8], and Varchenko, Ref.[9], in order to make a connection
between the P-F \ and R-G XXX\ spin chains. We use the results of our Part
II in order to demonstrate that the excitation spectrum of new Veneziano
model obtained in Part II coincides exactly with that for the R-G spin chain.

Since both ADS/CFT and this work point towards the same spin 1/2 XXX chain,
this cannot be considered as purely coincidental. Section 5 provides some
explanations of noticed coincidence based on \ independent combinatorial
arguments having their origin in the theory of random fragmentation and
coagulation processes summarized in Ref.s[38-40]. This theory was adopted
for high energy physics by Mekjian, e.g. see Ref.[41\textbf{]} and
references therein. A different approach to coagulation-fragmentation
processes in QCD is developed by Andersson and collaborators and is known in
literature as the Lund model, Ref.[42]. In this work no attempts are made to
compare these two approaches. Instead, in Section 5 we argue that in the
theory of coagulation-fragmentation processes, Veneziano amplitudes play the
central role. In this theory they are known as the Poisson-Diriclet (P-D)
probability distributions. The discrete spectra of all exactly solvable
quantum mechanical (QM), field and string-theoretic models can be
rederived/reobtained in terms of \ the observables for the P-D stochastic
processes. This is so because all exactly solvable QM problems involve some
kind of orthogonal polynomials-all derivable from the Gauss hypergeometric
function-admiting an interpretation in terms of the P-D process. Since the
K-Z equations are essentially the hypergeometric equations of many
variables, e.g. see Ref.[9], and since all \ nontrivial Feynman diagrams of
quantum field theory can be looked upon as solutions of these hypergeometric
equations as explained in Ref.[4], the processes they describe are also of
the P-D type.

Finally, Appendix contains several auxiliary results complementing those
presented in Sections 4 and 5.

\section{Combinatorics of Veneziano amplitudes and spin chains. Qualitative
considerations}

In Part I, we noticed that the Veneziano condition for the 4-particle
amplitude given by 
\begin{equation}
\alpha(s)+\alpha(t)+\alpha(u)=-1,  \tag{2.1}
\end{equation}
where $\alpha(s)$, $\alpha(t),\alpha(u)$ $\in\mathbf{Z}$, can be rewritten
in a more suggestive form. To this purpose, following Ref.[43], we need to
consider additional homogenous equation of the type 
\begin{equation}
\alpha(s)m+\alpha(t)n+\alpha(u)l+k\cdot1=0  \tag{2.2}
\end{equation}
with $m,n,l,k$ being some integers. By adding this equation to Eq.(2.1) we
obtain, 
\begin{equation}
\alpha(s)\tilde{m}+\alpha(t)\tilde{n}+\alpha(u)\tilde{l}=\tilde{k} 
\tag{2.3a}
\end{equation}
or, equivalently, as%
\begin{equation}
n_{1}+n_{2}+n_{3}=\hat{N},  \tag{2.3b}
\end{equation}
where all entries \textit{by design} are nonnegative integers. For the
multiparticle case this equation should be replaced by%
\begin{equation}
n_{0}+\cdot\cdot\cdot+n_{k}=N  \tag{2.4}
\end{equation}
so that \textsl{combinatorially the task lies in finding all nonnegative
integer combinations} \textsl{of} $n_{0},...,n_{k}$ \textsl{producing} \
Eq.(2.4). It should be noted that such a task makes sense as long as $N$ is
assigned. But the actual value of $N$ is \textit{not} \textit{fixed} and,
hence, can be chosen quite arbitrarily. Eq.(2.1) is a simple statement about
the energy-momentum conservation. Although the numerical entries in this
equation can be changed as we just explained, the actual physical values can
be \ subsequently reobtained by the appropriate coordinate shift. Such a
procedure should be applied to the amplitudes of conformal field theories
(CFT) \ with some caution \ since the periodic (or antiperiodic, etc.)
boundary conditions cause energy and momenta to become a \textit{quasi-energy%
} and a \textit{quasi momenta} (as \ it is known from solid state physics).

The arbitrariness of selecting $N$ reflects kind of a gauge freedom. As in
gauge theories, we may try to fix the gauge by using some physical
considerations. These include, for example, an observation made in Part I
that the four particle amplitude is zero if any two entries into Eq.(2.1)
are the same. This \ fact causes us to arrange the entries in Eq.(2.3b) in
accordance with their magnitudes, e.g. $n_{1}\geq n_{2}\geq n_{3}.$ More
generally, we can write: $n_{0}\geq n_{1}\geq\cdot\cdot\cdot\geq n_{k}\geq1%
\footnote{%
The last inequality: $n_{k}\geq1,$ is chosen only for the sake of comparison
with the existing literature conventions, e.g. see Ref.[44\textbf{]}.}$.

If the entries in this sequence of inequalities are treated as random
nonnegative numbers subject to the constraint given by Eq.(2.4), these
requirements are necessary and sufficient for recovery of the probability
density for such set of random numbers. This density \ is known in
mathematics as the Dirichlet distribution\footnote{%
For reasons explained in Section 5, it is also called the Poisson-Dirichlet
\ (P-D) distribution.} [38-40,45]. Without normalization, integrals over
this distribution coincide with Veneziano amplitudes. Details are given in
Section 5. \textsl{Thus, Veneziano condition \ leads to Veneziano amplitudes
and vice versa. In our work, }Ref.[4]\textsl{, we demonstrate that all
amplitudes of high energy physics are some linear} \textsl{combinations of
Veneziano amplitudes\footnote{%
This mathematical result was inspired by the observation that, at least
perturbatively, all these amplitudes are made of some linear combinations of
products of Euler gamma functions (with accuracy up to some logarithmic
terms).}}.

Provided that Eq.(2.4) holds, we shall call such a sequence a \textit{%
partition }and shall denote it as\textit{\ }$\mathit{n\equiv }%
(n_{0},...,n_{k})$. If $n$ is partition of $N$, then we shall write $n\vdash
N$. It is well known, e.g. see Ref.s[46,47], that there is one-to-one
correspondence between the Young diagrams and partitions. We would like to
use this fact in order to design a partition function associated with
Veneziano (and Veneziano-like) amplitudes\footnote{%
This task should not be confused with the task of connecting the P-D
distributions with Veneziano amplitudes to be discussed in Section 5.
Alternative pathway through methods of asymptotic combinatorics and
representation theory [48] will be treated in a separate publication.}.
Clearly, such a partition function should also make physical sense. Hence,
we would like to provide some qualitative arguments aimed at convincing our
readers that such a partition function does exist and is physically sensible.

We begin with observation that there is one-to-one correspondence between
the Young tableaux and directed random walks. Furthermore, it is possible to
map bijectively such type of random walk back into Young diagram with only
two rows, e.g. read Ref.[47], page 5. This allows us to make a connection
with spin chains at once. In this work we are not going to use this route to
spin chains in view of simplicity of \ the alternative path described in
this section\footnote{%
Nevertheless, this option should not be left underappreciated in view of its
immediate relevance to Hecke algebra representations, braid groups, etc.
e.g. see Ref.s[2] and[49].}. For this purpose we need to consider a square
lattice and to place on it the Young diagram associated with some particular
partition. Let us choose for this purpose some $\tilde{n}\times\tilde{m}$
rectangle\footnote{%
Parameters $\tilde{n}$ and $\tilde{m}$ will be specified shortly below.} so
that the Young diagram occupies the left part of this rectangle. We choose
the upper left vertex of the rectangle as the origin of $xy$ coordinate
system whose $y$ axis (South direction) is directed downwards and $x$ axis
is directed Eastwards. Then, the South-East boundary of the Young diagram
can be interpreted as directed (that is without self-intersections) random
walk which begins at $(0,-\tilde{m})$ and ends at $(\tilde{n},0).$
Evidently, such a walk completely determines the diagram. The walk can be
described by a sequence of 0's and 1's.\ Say, $0$ for the $x-$ step move and
1 for the $y-$step move. The totality $\mathcal{N}$ of Young diagrams which
can be placed into such a rectangle is in one-to-one correspondence with the
number of arrangements of 0's and 1's whose total number is $\tilde{m}+%
\tilde{n}$. Recalling the Fermi statistics, the number $\mathcal{N}$ can be
easily calculated and is given by $\mathcal{N}=(m+n)!/m!n!\footnote{%
We have suppressed the tildas for $n$ and $m$ in this expression since these
parameters are going to be redefined below anyway.}$. It can be represented
in two equivalent ways:%
\begin{align}
(m+n)!/m!n! & =\frac{(n+1)(n+2)\cdot\cdot\cdot(n+m)}{m!}\equiv\left( 
\begin{array}{c}
n+m \\ 
m%
\end{array}
\right)  \notag \\
& =\frac{(m+1)(m+2)\cdot\cdot\cdot(n+m)}{n!}\equiv\left( 
\begin{array}{c}
m+n \\ 
n%
\end{array}
\right) .  \tag{2.5}
\end{align}

Let now $p(N;k,m)$ be the number of partitions of $N$ into $\leq k$ \
nonnegative parts, each not larger than $m$. Consider the generating
function of the following type 
\begin{equation}
\mathcal{F}(k,m\mid q)=\dsum\limits_{N=0}^{S}p(N;k,m)q^{N},  \tag{2.6}
\end{equation}%
where the upper limit $S$\ will be determined shortly below. It is shown in
Refs.[44,46] that $\mathcal{F}(k,m\mid q)=\left[ 
\begin{array}{c}
k+m \\ 
m%
\end{array}%
\right] _{q}\equiv \left[ 
\begin{array}{c}
k+m \\ 
k%
\end{array}%
\right] _{q}$ where, for instance,$\left[ 
\begin{array}{c}
k+m \\ 
m%
\end{array}%
\right] _{q=1}=\left( 
\begin{array}{c}
k+m \\ 
m%
\end{array}%
\right) \footnote{%
On page 15 of the book by Stanley, Ref.[46], one can find that the number of
solutions $N(n,k)$ in \textit{positive} integers to $y_{1}+...+y_{k}=n+k$ is
given by $\left( 
\begin{array}{c}
n+k-1 \\ 
k-1%
\end{array}%
\right) $ while the number of solutions in \textit{nonnegative} integers to $%
x_{1}+...+x_{k}=n$ is $\left( 
\begin{array}{c}
n+k \\ 
k%
\end{array}%
\right) .$ Careful reading of Page 15 indicates however that the last number
refers to solution in nonnegative integers of the equation $x_{0}+...+x_{k}=n
$. This fact was used essentially in Eq.(1.21) of Part I.}.$ From this
result it should be clear that the expression $\left[ 
\begin{array}{c}
k+m \\ 
m%
\end{array}%
\right] _{q}$ is the $q-$analog of the binomial coefficient $\left( 
\begin{array}{c}
k+m \\ 
m%
\end{array}%
\right) .$ In literature [44,46] this $q-$ analog is known as the \textit{%
Gaussian} coefficient. Explicitly, it is defined as%
\begin{equation}
\left[ 
\begin{array}{c}
a \\ 
b%
\end{array}%
\right] _{q}=\frac{(q^{a}-1)(q^{a-1}-1)\cdot \cdot \cdot (q^{a-b+1}-1)}{%
(q^{b}-1)(q^{b-1}-1)\cdot \cdot \cdot (q-1)}  \tag{2.7}
\end{equation}%
for some nonegative integers $a$ and $b$. From this definition we anticipate
that the sum defining generating function $\mathcal{F}(k,m\mid q)$ in
Eq.(2.6) should have only \textit{finite} number of terms. Eq.(2.7) allows
easy determination of the upper limit $S$ in the sum given by Eq.(2.6). It
is given by $km$. This is just the area of the $k\times m$ rectangle. In
view of the definition of $p(N;k,m)$, the number $m=N-k$. Using this fact
Eq.(2.6) can be rewritten as $\mathcal{F}(N,k\mid q)=\left[ 
\begin{array}{c}
N \\ 
k%
\end{array}%
\right] _{q}.$This expression happens to be the Poincare$^{\prime }$
polynomial for the Grassmannian $Gr(m,k)$ of the complex vector space 
\textbf{C}$^{N}$of dimension $N$ as can be seen from page 292 of the book by
Bott and Tu, Ref.[50]\footnote{%
To make a comparison it is sufficient to replace parameters $t^{2}$ and $n$
in \ Bott and Tu book by $q$ and $N.$}. From this (topological) point of
view the numerical coefficients, i.e. $p(N;k,m),$ in the $q$ expansion of \
Eq.(2.6) should be interpreted as Betti numbers of this Grassmannian. They
can be determined recursively using the following property of the Gaussian
coefficientsdescribed in Ref.[46], page 26,%
\begin{equation}
\left[ 
\begin{array}{c}
n+1 \\ 
k+1%
\end{array}%
\right] _{q}=\left[ 
\begin{array}{c}
n \\ 
k+1%
\end{array}%
\right] _{q}+q^{n-k}\left[ 
\begin{array}{c}
n \\ 
k%
\end{array}%
\right] _{q},  \tag{2.8}
\end{equation}%
provided that $\left[ 
\begin{array}{c}
n \\ 
0%
\end{array}%
\right] _{q}=1.$ We refer our readers to Part II for rigorous mathematical
proof that $\mathcal{F}(N,k\mid q)$ is indeed the Poincare$^{\prime }$
polynomial for the complex Grassmannian. With this fact proven, we notice
that, due to relation $m=N-k,$ it is sometimes more convenient \ for us to
use the parameters $m$ and $k$ rather than $N$ and $k$. With such a
replacement we obtain: 
\begin{align}
\mathcal{F}(k,m& \mid q)=\left[ 
\begin{array}{c}
k+m \\ 
k%
\end{array}%
\right] _{q}=\frac{(q^{k+m}-1)(q^{k+m-1}-1)\cdot \cdot \cdot (q^{m+1}-1)}{%
(q^{k}-1)(q^{k-1}-1)\cdot \cdot \cdot (q-1)}  \notag \\
& =\dprod\limits_{i=1}^{k}\frac{1-q^{m+i}}{1-q^{i}}.  \tag{2.9}
\end{align}%
This result is of central importance. In our work, Part II, considerably
more sophisticated mathematical apparatus was used to obtain it (e.g. see
Eq.(6.10) of this reference and arguments leading to it).\ 

In the limit $q\rightarrow 1$ Eq.(2.9) reduces to $\mathcal{N}$ as required.
To make connections with results known in physics literature we need to
rescale $q^{\prime }s$ in \ Eq.(2.9), e.g. let $q=t^{\frac{1}{i}}.$
Substitution of such an expression back into Eq.(2.9) and taking the limit $%
t\rightarrow 1$ again reproduces $\mathcal{N}$ in view of Eq.(2.5). This
time, however, we can accomplish more. By noticing that in Eq.(2.4) the
actual value of $N$ deliberately is not yet fixed and taking into account
that $m=N-k,$ we can fix $N$ by fixing $m$. Specifically, we would like to
choose $m=1\cdot 2\cdot 3\cdot \cdot \cdot k$ and with such a choice we
would like\ to consider a particular term in the product, Eq.(2.9), e.g.%
\begin{equation}
S(i)=\frac{1-t^{1+\frac{m}{i}}}{1-t}.  \tag{2.10}
\end{equation}%
In view of our "gauge fixing" the ratio $m/i$ is a positive integer by
design. This means that we are having a geometric progression. Indeed, if we
rescale $t$ again , e.g. $t\rightarrow t^{2},$ we then obtain%
\begin{equation}
S(i)=1+t^{2}+\cdot \cdot \cdot +t^{2\hat{m}}  \tag{2.11}
\end{equation}%
with $\hat{m}=\frac{m}{i}.$ Written in such a form the \ above sum is just
the Poincare$^{\prime }$ polynomial for the complex projective space \textbf{%
CP}$^{\hat{m}}.$ This can be seen by comparing pages 177 and 269 of the book
by Bott and Tu, Ref.[50]. Hence, at least for some $m$'s, \textit{the
Poincare}$^{\prime }$\textit{\ polynomial for the Grassmannian in just the
product of} \textit{the }Poincare$^{\prime }$\textit{\ polynomials for the
complex projective spaces of known dimensionalities}. For $m$ just chosen,
in the limit $t\rightarrow 1,$ we reobtain back the number $\mathcal{N}$ as
required. This physically motivating process of gauge fixing \ we just
described will be replaced by more rigorous mathematical arguments in the
rest of this paper. Rigorous mathematical arguments causing factorization of
the Poincare$^{\prime }$ polynomial can be found, for instance, in Ch-3 of
lecture notes by Schwartz, Ref.[51]. The relevant physics emerges by
noticing that the partition function $Z(J)$ for the particle with spin $J$
is given by, e.g. see Ref.[52], 
\begin{align}
Z(J)& =tr(e^{-\beta H(\sigma )})=e^{cJ}+e^{c(J-1)}+\cdot \cdot \cdot +e^{-cJ}
\notag \\
& =e^{cJ}(1+e^{-c}+e^{-2c}+\cdot \cdot \cdot +e^{-2cJ}),  \tag{2.12}
\end{align}%
where $c$ is known constant. Evidently, up to a constant, $Z(J)\simeq S(i).$
Since\ mathematically the result, Eq.(2.12), is the Weyl character formula,
this fact brings the classical group theory into our discussion. More
importantly, because the partition function for the particle with spin $J$
can be written in the language of N=2 supersymmetric quantum mechanical model%
\footnote{%
We hope that no confusion is made about the meaning of N in the present case.%
}, as demonstrated by Stone, Ref.[52] and others, Ref.[53], the connection
between the supersymmetry and the classical group theory is evident. \ It
was developed to a some extent in Part III.

In view of arguments presented above, the Poincare$^{\prime}$ polynomial for
the Grassmannian can be interpreted as a partition function for some kind of
a spin chain made of \ apparently independent spins of various magnitudes%
\footnote{%
In such a context it can be vaguely considered as a variation on the theme
of the Polyakov rigid string (Grassmann $\sigma$ model, Ref.[54], pages
283-287), except that now it is \textit{exactly solvable} in the qualitative
context \ just described and, below, in mathematically rigorous context.}.
These qualitative arguments we would like to make more mathematically and
physically rigorous. The first step towards this goal is made in the next
section.

\section{Connection with the Polychronakos-Frahm spin chain model}

\bigskip

The Polychronakos-Frahm (P-F) spin chain model \ was originally proposed by
Polychronakos and described in detail in Ref.[55]. Frahm [56] \ motivated by
the results of Polychronakos made additional progress in elucidating the
spectrum and thermodynamic properties of this model so that it had become
known as the P-F model. Subsequently, many other researchers have
contributed to our understanding of this exactly integrable spin chain
model. Since this paper is not a review, we shall quote only those works on
P-F model which are of immediate relevance.

Following \ Ref.[55], we begin with some description of the P-F model. Let $%
\sigma _{i}^{a}$ ($a=1,2,...,n^{2}-1)$ be $SU(n)$ spin operator of i-th
particle and let the operator$\ \sigma _{ij}$ be responsible for a spin
exchange between particles $i$ and $j,$ i.e.%
\begin{equation}
\sigma _{ij}=\frac{1}{n}+\tsum\limits_{a}\sigma _{i}^{a}\sigma _{j}^{a}. 
\tag{3.1}
\end{equation}%
In terms of these definitions, the Calogero-type model Hamiltonian can be
written as [57,58] 
\begin{equation}
\mathcal{H}=\frac{1}{2}\tsum\limits_{i}(p_{i}^{2}+\omega
^{2}x_{i}^{2})+\tsum\limits_{i<j}\frac{l(l-\sigma _{ij})}{\left(
x_{i}-x_{j}\right) ^{2}},  \tag{3.2}
\end{equation}%
where $l$ is some parameter. The P-F model is obtained from the above model
in the limit $l\rightarrow \pm \infty $ . Upon proper rescaling of $\mathcal{%
H}$ in Eq.(3.2), in this limit one obtains:%
\begin{equation}
\mathcal{H}_{\mathcal{P-F}}=-sign(l)\tsum\limits_{i<j}\frac{\sigma _{ij}}{%
\left( x_{i}-x_{j}\right) ^{2}},  \tag{3.3}
\end{equation}%
where the coordinate $x_{i}$ minimizes the potential for the rescaled
Calogero model\footnote{%
The Calogero model is obtainable from the Hamiltonian, Eq.(3.2), if one
replaces the spin exchange operator $\sigma _{ij}$ by 1. Since we are
interested in the large $l$ limit, one can replace the factor $l(l-1)$ by $%
l^{2}$ in the interaction term.}, that is%
\begin{equation}
\omega ^{2}x_{i}^{{}}=\tsum\limits_{i<j}\frac{2}{\left( x_{i}-x_{j}\right)
^{3}}.  \tag{3.4}
\end{equation}%
It should be noted that $\mathcal{H}_{\mathcal{P-F}}$ is well defined
without such a minimization, that is for arbitrary real parameters $x_{i}$.
This fact will be further explained in Section 4.\ In the large $l$ limit
the spectrum of $\mathcal{H}$ is decomposable as 
\begin{equation}
E=E_{\mathcal{C}}+lE_{\mathcal{P-F}},  \tag{3.5}
\end{equation}%
where $E_{C}$ is the spectrum of spinless Calogero model while $E_{\mathcal{%
P-F}}$ is the spectrum of P-F model. In view of such a decomposition, the
partition function for the Hamiltonian $\mathcal{H}$ at temperature $T$ can
be written as a product: Z$_{\mathcal{H}}(T)=$Z$_{\mathcal{C}}(T)$Z$_{%
\mathcal{P-F}}(T/l)$. From here, one formally obtains the result: 
\begin{equation}
Z_{\mathcal{P-F}}(T)=\lim_{l\rightarrow \infty }\frac{Z_{\mathcal{H}}(lT)}{%
Z_{\mathcal{C}}(T)}.  \tag{3.6}
\end{equation}%
It implies that the spectrum of P-F spin chain can be obtained if both the
total and \ the Calogero partition functions can be calculated. In Ref.[55]
Polychronakos argued that $Z_{\mathcal{C}}(T)$ is essentially a partition
function of $\mathit{N}$ noninteracting harmonic oscillators. Thus, we
obtain: 
\begin{equation}
Z_{\mathcal{C}}(N;T)=\tprod\limits_{i=1}^{N}\frac{1}{1-q^{i}},\text{ }q=\exp
(-\beta ),\beta =\left( k_{B}T\right) ^{-1}.  \tag{3.7}
\end{equation}%
Furthermore, the partition function $Z_{\mathcal{H}}(T)$ according to
Polychronakos can be obtained using $Z_{\mathcal{C}}(N;T)$ as follows.
Consider the grand partition function of the type%
\begin{equation}
\Xi =\tsum\limits_{N=0}^{\infty }Z_{n}(N;T)y^{N}\equiv \left(
\tsum\limits_{L=0}^{\infty }Z_{\mathcal{C}}(L;T)y^{L}\right) ^{n},  \tag{3.8}
\end{equation}%
where $n$ is the number of flavors\footnote{%
That is $n$ the same number as $n$ in $SU(n).$}. \ Using this definition we
obtain:%
\begin{equation}
Z_{n}(N;T)=\sum_{\Sigma _{i}k_{i}=N}\prod\limits_{i=1}^{n}Z_{\mathcal{C}%
}(k_{i};T).  \tag{3.9}
\end{equation}%
Next, Polychronakos identifies $Z_{n}(N;T)$ with Z$_{\mathcal{H}}(T)$. Then,
with help of \ Eq.(3.6) the partition function $Z_{\mathcal{P-F}}(T)$ is
obtained straightforwardly as 
\begin{equation}
Z_{\mathcal{P-F}}(N;T)=\sum_{\Sigma _{i}k_{i}=N}\frac{\tprod%
\limits_{i=1}^{N}(1-q^{i})}{\prod\limits_{i=1}^{n}\prod%
\limits_{r=1}^{k_{i}^{{}}}(1-q^{r})}.  \tag{3.10}
\end{equation}%
Consider this result for a special case $n=2$. It is convenient to evaluate
the ratio first before calculating the sum. Thus, we obtain:%
\begin{equation}
\frac{\tprod\limits_{i=1}^{N}(1-q^{i})}{\prod\limits_{i=1}^{2}\prod%
\limits_{r=1}^{k_{i}^{{}}}(1-q^{r})}=\frac{(1-q)\cdot \cdot \cdot (1-q^{N})}{%
(1-q)\cdot \cdot \cdot (1-q^{k})(1-q)\cdot \cdot \cdot (1-q^{N-k})}\equiv 
\mathcal{F}(N,k\mid q),  \tag{3.11}
\end{equation}%
where the Poincare$^{\prime }$ polynomial $\mathcal{F}(N,k\mid q)$ for the
Grassmanian of the complex vector space \textbf{C}$^{N}$ of dimension $N$ \
was obtained in the previous section. Indeed, \ Eq.(3.11) can be trivially
brought into the same form as that given in our Eq.(2.9) using the relation $%
m+k=N$. To bring\ our Eq.(2.9) in correspondence with Eq.(4.1) of
Polychronakos, Ref.[55], we use the second \ line in Eq.(2.9) \ in which we
make a substitution: $m=N-k.$ After this replacement, Eq.(3.10) acquires the
form%
\begin{equation}
Z_{\mathcal{P-F}}^{f}(N;T)=\sum\limits_{k=0}^{N}\prod\limits_{i=0}^{k}\frac{%
1-q^{N-i+1}}{1-q^{i}}  \tag{3.12}
\end{equation}%
coinciding with Eq.(4.1) by Polychronakos. This equation corresponds to the
ferromagnetic version of the P-F spin chain model. To obtain the
antiferromagnetic version of the model requires us only to replace $q$ by $%
q^{-1}$ in \ Eq.(3.12) and to multiply the whole r.h.s. by some known power
of $q$. Since this factor will not affect thermodynamics, following Frahm,
Ref.[56], we shall ignore it. As result, we obtain 
\begin{equation}
Z_{\mathcal{P-F}}^{af}(N;T)=\sum\limits_{k=0}^{N}q^{\left( N/2-k\right)
^{2}}\prod\limits_{i=0}^{k}\frac{1-q^{N-i+1}}{1-q^{i}},  \tag{3.13}
\end{equation}%
in accord with Frahm's \ Eq.(21). \ This result is analyzed further in the
next section.

\section{Connections with WZNW model and XXX \ s=1/2 \ Heisenberg
antiferromagnetic \ spin chains}

\subsection{General remarks}

\bigskip To establish these connections we follow work by Hikami, Ref.[59].
For this purpose, we introduce the notation%
\begin{equation}
\left( q\right) _{n}=\prod\limits_{i=1}^{n}(1-q^{i})  \tag{4.1}
\end{equation}
allowing us to rewrite \ Eq.(3.13) in the equivalent form%
\begin{equation}
Z_{\mathcal{P-F}}^{af}(N;T)=\sum\limits_{k=0}^{N}q^{\left( N/2-k\right)
^{2}}\prod\limits_{i=0}^{k}\frac{1-q^{N-i+1}}{1-q^{i}}=\sum%
\limits_{k=0}^{N}q^{\left( N/2-k\right) ^{2}}\frac{\left( q\right) _{N}}{%
\left( q\right) _{k}\left( q\right) _{N-k}}.  \tag{4.2}
\end{equation}
Consider now the limiting case ($N\rightarrow\infty)$ of the obtained
expression. \ For this purpose we need to take into account that 
\begin{equation}
\lim_{N\rightarrow\infty}\left[ 
\begin{array}{c}
N \\ 
k%
\end{array}
\right] _{q}=\frac{1}{\left( q\right) _{k}}.  \tag{4.3}
\end{equation}
To use this asymptotic result in \ Eq.(4.2) it is convenient to consider
separately the cases of $N$ being even and odd. For instance, if $N$ is
even, we can write: $N=2m.$ In such a case we can introduce new summation
variables: $l=k-m$ and/or $l=m-k.$ Then, in the limit $N\rightarrow\infty$
(that is if m$\rightarrow\infty)$ we obtain asymptotically%
\begin{equation}
Z_{\mathcal{P-F}}^{af}(\infty;T)=\frac{1}{\left( q\right) _{\infty}}%
\sum\limits_{i=-\infty}^{\infty}q^{i^{2}}.  \tag{4.4a}
\end{equation}
in accord with Ref.[59]. Analogously, if $N=2m+1$, we obtain instead 
\begin{equation}
Z_{\mathcal{P-F}}^{af}(\infty;T)=\frac{1}{\left( q\right) _{\infty}}%
\sum\limits_{i=-\infty}^{\infty}q^{\left( i+\frac{1}{2}\right) ^{2}}. 
\tag{4.4b}
\end{equation}
According to Melzer, Ref.[60], and Kedem, McCoy and Melzer, Ref.[61], the
obtained partition functions coincide with the Virasoro characters for SU$%
_{1}$(2) WZNW \ model describing the conformal limit of the XXX (s=1/2)
antiferromagnetic spin chain, e.g. see Ref.[62]. Even though \ Eq.s(4.4a)
and (4.4b) provide the final result, they do not reveal their physical
content. This task was accomplished in part in the same papers where
connection with the excitation spectrum of the XXX antiferromagnetic chain \
was made. Hence, at the physical level of rigor\ the problem of connecting
Veneziano amplitudes with physical model can be considered as solved.
Nevertheless, below we argue that at the mathematical level of rigor this is
not quite so yet. This conclusion concerns not only problems dicussed in
this paper but, in general, the connection between the WZNW models, spin
chains and K-Z equations.

It is true that K-Z equations and WZNW model \ are inseparable from each
other as explained, for example, in Ref.[62] but the extent\ to which spin
chains can be directly linked to both \ the WZNW models and K-Z equations
still remains to be discussed. For the sake of space, we shall discuss only
the most essential facts leaving (with few exceptions) many details and
proofs to literature.

Following Varchenko, Ref.[9], we notice that the link between the K-Z
equations and WZNW models can be made\ only with help of\ the Gaudin model,
while the connection with spin chains can be made only by using the \textsl{%
quantum} version of K-Z equations. Such quantized version of K-Z equations 
\textsl{is not} immediately connected with the standard WZNW model \ as
discussed in many places, e.g. see Ref.s[9,63]. Therefore, we would like to
discuss in some detail the Gaudin model \ first and only then its relation
to P-F spin chain and, accordingly, with the Veneziano model formulated \
and studied in Part II.

\subsection{Gaudin magnets, K-Z equation and P-F spin chains}

Although theory of the Gaudin magnets plays an important role in topics such
as Langlands correspondence, Hitchin systems, etc., as explained, for
instance, in Ref.s[64-66], in this work we do not discuss these topics.
Instead, we would like to focus only on issues of \ immediate relevance to
this paper.\ Gaudin came up with his magnetic chain model in 1976, Ref.[67],
being influenced by earlier works of Richardson, Ref.s[68,69] on exact
solution of the BCS equations of superconductivity. This connection \ with
superconductivity will play an important role in what follows.

In physics literature\ all Gaudin-type models are based on $SU(2)$ algebra
of spin operators\footnote{%
In mathematics literature to be used below [9,63] the $SL(2,C)$ group is
used instead of \ its subgroup, $SU(2)$ [70].}. Instead of one Hamiltonian,
the set of commuting Hamiltonians of the type%
\begin{equation}
H_{i}=\sum\limits_{j(\neq i)=1}^{N}\sum\limits_{\alpha =1}^{3}w_{ij}^{\alpha
}\sigma _{i}^{\alpha }\sigma _{j}^{\alpha }  \tag{4.5}
\end{equation}
is used as discussed in Ref.[71]. In view of the fact that, by construction, 
$[H_{i},H_{j}]=0$, and$3N(N-1),$ the coefficients $w_{ij}^{\alpha }$ should
satisfy the following equations%
\begin{equation}
w_{ij}^{\alpha }w_{jk}^{\gamma }+w_{ji}^{\beta }w_{ik}^{\gamma
}-w_{ik}^{\alpha }w_{jk}^{\beta }=0.  \tag{4.6}
\end{equation}%
These equations can be solved by imposing the antisymmetry requirement: $%
w_{ij}^{\alpha }=-w_{ji}^{\alpha }$. It can be satisfied by replacing $%
w_{ij}^{\alpha }$ by the unknown functions $w_{ij}^{\alpha }=f^{\alpha
}(z_{i}-z_{j})$ of \ difference between two new real parameters $z_{i}$ and $%
z_{j}$. It is only natural to make further restrictions based on requirement
that the $z-$component of the total spin $S^{3}$ =$\sum\nolimits_{i}\sigma
_{i}^{3}$ is conserved. This causes us to write $w_{ij}^{1}=w_{ij}^{2}\equiv
X_{ij},$ and $w_{ij}^{3}=Y_{ij}$ thus leading to equations 
\begin{equation}
Y_{ij}X_{jk}+Y_{ki}X_{jk}+X_{ki}X_{ij}=0.  \tag{4.7}
\end{equation}%
These constraint equations admit the following sets of solutions: 
\begin{equation}
X_{ij}=Y_{ij}=\frac{1}{z_{i}-z_{j}}\text{ \ \ (rational),}  \tag{4.8a}
\end{equation}%
\begin{equation}
X_{ij}=\frac{1}{\sin \left( z_{i}-z_{j}\right) }\text{ , }Y_{ij}=\cos \left(
z_{i}-z_{j}\right) \text{ \ (trigonometric),}  \tag{4.8b}
\end{equation}%
\begin{equation}
X_{ij}=\frac{1}{\sinh \left( z_{i}-z_{j}\right) },\text{ }Y_{ij}=\cosh
\left( z_{i}-z_{j}\right) \text{ (hyperbolic).}  \tag{4.8c}
\end{equation}%
While the first solution, Eq.(4.8a), to be used in this work, \ corresponds
to the long \ range analog of the standard $XXX$ spin chain, the remaining
two solutions correspond to the long range analogs of the $XXZ$ spin chain.

Folloving Varchenko, Ref.[9], we are now in the position to write down the
K-Z equations. For this purpose we combine Eq.s (4.5) and (4.8a) \ and
reintroduce the coupling constant $g$ (so that $w_{ij}^{\alpha }\rightarrow
gw_{ij}^{\alpha })$ in such a way that the set of K-Z equations acquires the
form%
\begin{equation}
(\kappa \frac{\partial }{\partial z_{i}}-H_{i}(z_{1},...,\text{ }z_{N}))\Phi
(z_{1},...,\text{ }z_{N})=0,\text{ \ \ \ \ }i=1,_{...},N,  \tag{4.9}
\end{equation}%
where $\kappa =g^{-1}.$ This result requires several comments. First, from
theory of WZNW models it is known that parameter $\kappa $ cannot take
arbitrary values. For instance, for $SU_{1}(2)$ WZNW model $\kappa =\frac{3}{%
2},$ e.g. read Ref.[62]. Second, we can always rescale $z$-coordinates and
to redefine the Hamiltonian to make the constant arbitrary small.
Apparently, this \ is asssumed in the asymptotic analysis of K-Z equations
described in Ref.s[8,9]. Third, if this is the case, then such analysis (to
be used below) differs essentially from other approaches connecting string
models with spin chains discussed in the Introduction since such a
connection \ was typically made in the limit $N\rightarrow \infty $. Since
for $SU(N)$-type models $\kappa =\frac{1}{2}(k+N),$ in the limit $%
N\rightarrow \infty $ we have $\kappa \rightarrow \infty .$ The WKB-type
analysis of K-Z equations of Reshetikhin and Varchenko (to be discussed
below) fails exactly in this limit.

With set of K-Z equations defined, we would like now to make a connection
between the Gaudin and P-F model. To a large extent this was already
accomplished in Ref.[72]. Following this reference, we define the spin
Calogero (S-C) model as follows 
\begin{equation}
H_{S-C}=\frac{1}{2}\sum\limits_{i}^{{}}(p_{i}^{2}+\omega
^{2}x_{i}^{2})+g\sum\limits_{\substack{ i<j  \\ }}^{{}}\frac{\vec{\sigma}%
_{i}\cdot \vec{\sigma}_{j}}{\left( z_{i}-z_{j}\right) ^{2}}  \tag{4.10}
\end{equation}%
to be compared with $\mathcal{H}$ in \ Eq.(3.2)\footnote{%
We added the oscillator-type potential absent in the original work,
Ref.[72], for the sake of additional comparisons, e.g. with \ Eq.(3.4). In
what follows such a constraint is not essential and will be ignored.}. Using
the rational form of the Gaudin Hamiltonian this result can be equivalently
rewritten as 
\begin{equation}
H_{S-C}=\frac{1}{2}\sum\limits_{l}^{{}}(p_{l}^{2}+\omega ^{2}x_{l}^{2}+i%
\frac{g}{2}[p_{l},H_{l}]).  \tag{4.11}
\end{equation}%
That this is indeed the case can be seen by the following chain of arguments.

Consider the strong coupling limit ($g\rightarrow\infty)$ of $H_{S-C}$ so
that the kinetic term is a perturbation. Next, we consider the eigenvalue
problem for one of the Gaudin's Hamiltonians, i.e.%
\begin{equation}
H_{l}\Psi^{(l)}=E^{(l)}\Psi^{(l)},  \tag{4.12}
\end{equation}
and apply the operator $ip_{l}$ to both sides of this equation. Furthermore,
consider in \textsl{this} limit the combination $H_{S-C}\Psi^{(l)}.$
Provided that the eigenvalue problem, Eq.(4.12), does have a solution, it is
always possible to Fourier expand ($ip_{l}\Psi^{(l)})$ using as basis set $%
\Psi ^{(l)}.$ In such a case we end up with the eigenvalue problem for the
P-F spin chain in which the eigenfunctions are the same as those for the
Gaudin's model and the eigenvalues are $ip_{l}E^{(l)}.$ Physical
significance of this result \ will be discussed in detail below. Before
doing so, we have to make a connection between the K-Z, \ Eq.(4.9), and the
Gaudin eigenvalue, Eq.(4.12),\ problems.

Following Ref.s[8,9], we begin by replacing $SU(2)$ spin operators by the $%
SL(2,C)\equiv sl_{2}$ operators $e$, $f$ and $h$ obeying commutation
relations 
\begin{equation}
\lbrack h,e]=2e;\text{ \ \ }[e,f]=h;\text{ \ }[h,f]=-2f.  \tag{4.13}
\end{equation}
This \ Lie algebra was discussed extensively in Part II in connection with \
design of new models reproducing Veneziano amplitudes. \ In this work, we
shall extend already obtained results following \ ideas of Richardson and
Varchenko.

From Ref.[70] it is known \ that $SU(2)$ is just a subgroup of $sl_{2}.$
Introduce the Casimir element $\Omega$ $\in$ $sl_{2}\otimes sl_{2}$ via%
\begin{equation}
\Omega=e\otimes f+f\otimes e+\frac{1}{2}h\otimes h  \tag{4.14}
\end{equation}
so that $\forall x\in sl_{2}$ it satisfies the commutation relation $%
[x\otimes1+1\otimes x,$ $\Omega]=0$ inside $\ $the $U(sl_{2})\otimes
U(sl_{2}),$ where $U(sl_{2})$ is the universal enveloping algebra of $%
sl_{2}. $ Consider the vector space $V=V_{1}\otimes
V_{2}\otimes\cdot\cdot\cdot\otimes V_{N}$. An element $x\in sl_{2}$ acts on $%
V$ as follows: $x\otimes
1\otimes1\otimes\cdot\cdot\cdot\otimes1+\cdot\cdot\cdot+1\otimes1\otimes
\cdot\cdot\cdot\otimes x.$ For indices $1\leq i<j\leq N $ \ let $\Omega
^{(i,j)}:V\rightarrow V$ be an operator which acts as $\Omega$ on i-th and
j-th positions and as identity on all others, then the set of K-Z equations
can be written as 
\begin{equation}
\kappa\frac{\partial}{\partial z_{i}}\Phi=\sum\limits_{j\neq i}\frac {%
\Omega^{(i,j)}}{z_{i}-z_{j}}\Phi\text{, }\ i=1,...,N.  \tag{4.15}
\end{equation}
In the simplest case, this set of equations is defined in the domain $%
U=\{(z_{1},...,z_{N})\in\mathbf{C}^{N}\mid z_{i}\neq z_{j}\}.$

From \ now on we shall use Eq.s(4.15) instead of \ Eq.s(4.9). To connect K-Z
equations with the XXX Gaudin magnet we shall use the WKB method developed
by Reshetikhin and Varchenko, Ref.[8], and summarized in lecture notes by
Varchenko, Ref.[9]. Following these authors, we \ shall look for a solution
of Eq.(4.15) in the form ($\kappa \rightarrow 0)$%
\begin{equation}
\Phi (\mathbf{z},\kappa )=e^{\frac{1}{\kappa }S(\mathbf{z})}\{f_{0}(\mathbf{z%
})+\kappa f_{1}(\mathbf{z})+\cdot \cdot \cdot \},  \tag{4.16}
\end{equation}%
where $\mathbf{z}=\{z_{1},...,z_{N}\}$, $S(\mathbf{z})$ is some scalar
function (to be described below) and $f_{j}(\mathbf{z})$, $j=0,1,2,...$, are 
$V-$valued functions. Provided that the function $S$ is known, $V-$valued
functions can be recursively determined (as it is done in the WKB analysis).
Specifically, given that $H_{i}=\sum\limits_{j\neq i}\dfrac{\Omega ^{(i,j)}}{%
z_{i}-z_{j}},$ we obtain:%
\begin{equation}
H_{i}f_{0}(\mathbf{z})=\frac{\partial S}{\partial z_{i}}f_{0}(\mathbf{z}), 
\tag{4.17}
\end{equation}%
to be compared with \ Eq.(4.12). Next, we get 
\begin{equation}
H_{i}f_{1}(\mathbf{z})=\frac{\partial S}{\partial z_{i}}f_{1}(\mathbf{z})+%
\frac{\partial f_{0}}{\partial z_{i}},  \tag{4.18}
\end{equation}%
and so on. \ Since the function $S(\mathbf{z})$ (the Shapovalov form) plays
\ an important role in these calculations, we would like to discuss it in
some detail now.

\subsection{The Shapovalov form}

\bigskip

Consider the following auxiliary problem. Let $A(x)$ and $B(x)$ be some pre
assigned polynomials of degree $n$ and $n-1$ respectively. Find a polynomial 
$C(x)$ of degree $n-2$ such that the differential equation%
\begin{equation}
A(x)y^{^{\prime\prime}}(x)-B(x)y^{\prime}(x)+C(x)y(x)=0  \tag{4.19}
\end{equation}
has solution which is polynomial of preassigned degree $k$. Such polynomial
solution is called the Lame$^{\prime}$ function. Stieltjes, Ref.s[8,9],
proved the following

\medskip

\textbf{Theorem 4.1}. \textsl{Let} $A$ \textsl{and} $B$ \textsl{be given
polynomials of degree} $n$ \textsl{and} $n-1$, \textsl{respectively so that} 
$B(x)/A(x)=\sum\nolimits_{j=1}^{n}\dfrac{m_{j}}{x-x_{j}}$. \textsl{Then
there is a polynomial} $C$ \textsl{of degree} $n-2$ \textsl{and a polynomial
solution} $y(x)=\prod\nolimits_{i=1}^{k}(x-x_{i})$ \textsl{of} \ Eq.(4.19) 
\textsl{if and only if} $\mathbf{\check{x}=}(x_{1},...,x_{k})$ \textsl{is
the critical point of the function}%
\begin{equation}
\Phi_{k,n}(x_{1},...,x_{k};z_{1},...,z_{n})=\prod\limits_{j=1}^{k}\prod%
\limits_{i=1}^{n}(x_{j}-z_{i})^{-m_{i}}\prod\limits_{1\leq i<j\leq
k}(x_{i}-x_{j})^{2}.  \tag{4.20}
\end{equation}

\medskip

\textbf{Definition 4.2}. A point $\mathbf{\check{x}}$\textbf{\ }is \textsl{%
critical} for\textbf{\ }$\Phi(\mathbf{x})$ if all its first derivatives
vanish at it.

\quad\ 

We would like now to make a connection between the Shapovalov form $S$ and
results just obtained. $S$ is symmetric bilinear form on previously
introduced space $V$ such that $S(v,v)=1$, $S(hx,y)=S(x,hy),$ $%
S(ex,y)=S(x,fy),$ where $h,e,f$ are defined in Eq.(4.13). Furthermore, $%
S(\Omega (x_{1}\otimes x_{2}),y_{1}\otimes y_{2})=S(x_{1}\otimes
x_{2},\Omega (y_{1}\otimes y_{2}))$ $\forall x_{1},y_{1}\in V_{1}$ and $%
\forall x_{2},y_{2}\in V_{2}.$ As result, we obtain:%
\begin{equation}
S(H_{i}x,y)=S(x,H_{i}y)\text{ }\forall x,y\in V.  \tag{4.21}
\end{equation}%
Next, let $m$ be some nonnegative integer and $V_{m}$ be the irreducible\
Verma module with the highest weight $m$ and the highest weight singular
vector $v_{m},$ i.e.%
\begin{equation}
hv_{m}=mv_{m},\text{ }ev_{m}=0.  \tag{4.22}
\end{equation}%
Consider a tensor product $V\equiv V^{\otimes M}=V_{m_{1}}\otimes \cdot
\cdot \cdot \otimes V_{m_{n}}$ so that $M=(m_{1},...,m_{n}).$ $\forall
V_{m_{i}}$ vectors $v_{m_{i}}$, $fv_{m_{i}},f^{2}v_{m_{i}},\cdot \cdot \cdot
,f^{m_{i}}v_{m_{i}}$ form a basis of $V_{m_{i}}$ \footnote{%
According to Ref.[9] in all subsequent calculations it is sufficient to use%
\textsl{\ the finite} Verma module, i.e. $L_{m}=V_{m}/<f^{m+1}v_{m}>.$ This
restriction is in accord \ with our previous calculations, e.g. see Part II,
Section 8, where such a restriction originates from the Lefschetz
isomorphism theorem used in conjunction with supersymmetric model
reproducing Veneziano amplitudes.}so that the Shapovalov form is orthogonal
with respect to such a basis and is decomposable as $S=S_{m_{1}}\otimes
\cdot \cdot \cdot \otimes S_{m_{n}}$ \ Let, furthermore, $%
J=(j_{1},...,j_{n}) $ be a set of nonnegative integers such that $%
j_{1}+\cdot \cdot \cdot +j_{n}=k,$ where $k$ is the same as in Eq.(4.20),
and $0\leq j_{i}\leq m_{l}$. This allows us to define the set of vectors $%
f^{J}v_{M}=f^{j_{1}}v_{m_{1}}\otimes \cdot \cdot \cdot \otimes
f^{j_{n}}v_{m_{n}}$ . \ These vectors $\{f^{J}v_{M}\}$ are by construction
orthogonal with respect to the Shapovalov form and provide a basis for the
space $V^{\otimes M}.$ Introduce the \textsl{weight} of a partition $A$ as $%
\left\vert A\right\vert =a_{1}+a_{2}+...$ then, in view of Eq.(4.22), we
define the singular vector $f^{J}v_{M}$ via 
\begin{equation}
h(f^{J}v_{M})=(\left\vert M\right\vert -2\left\vert J\right\vert )f^{J}v_{M}%
\text{ , \ \ }e\text{(}f^{J}v_{M})=0  \tag{4.23}
\end{equation}%
of weight $\left\vert M\right\vert -2\left\vert J\right\vert .\footnote{%
This fact can be easily undestood from the properties of $sl_{2}$ Lie
algebra representations since it is known, e.g see Ref.[9] and Part II, that
for the module of highest weight $m$ we have $%
h(f^{k}v_{m})=(m-2k)(f^{k}v_{m}).$}$ The Bethe ansatz vectors $\mathcal{V}$
for the Gaudin model can be defined now as 
\begin{equation}
\mathcal{V(}\mathbf{\check{x}},\mathbf{z})=\sum\limits_{J}A_{J}(\mathbf{%
\check{x}},\mathbf{z})f^{J}v_{M},  \tag{4.24}
\end{equation}%
where \textbf{\v{x}} \ is a critical point of $\Phi (\mathbf{x,z})$ defined
by Eq.(4.20) while the function $A_{J}(\mathbf{\check{x}},\mathbf{t})$ is
defined as follows%
\begin{equation}
A_{J}(\mathbf{\check{x}},\mathbf{z})=\sum\limits_{\sigma \in \mathcal{P(}%
k;J)}\prod\limits_{i=1}^{k}\frac{1}{x_{i}-z_{\sigma (i)}}  \tag{4.25}
\end{equation}%
with $\mathcal{P(}k;J)$ being the set of maps $\sigma $ from the $%
\{1,...,k\} $ to $\{1,...,n\}$. Finally, using these definitions it is
possible to prove that%
\begin{equation}
S(\mathcal{V(}\mathbf{\check{x}},\mathbf{z}),\mathcal{V(}\mathbf{\check{x}},%
\mathbf{z}))=\det_{1\leq i,j\leq k}(\frac{\partial ^{2}}{\partial
x_{i}\partial x_{j}}\ln \Phi _{k,n}(\check{x}_{1},...,\check{x}%
_{k};z_{1},...,z_{n})).  \tag{4.26}
\end{equation}%
The set of equations determining critical points 
\begin{equation}
\frac{1}{\Phi _{k,n}(\mathbf{x}^{0},\text{ }\mathbf{z}^{0})}\frac{\partial }{%
\partial x_{i}}\Phi _{k,n}(\mathbf{x}(\mathbf{z}),\text{ }\mathbf{z})\mid _{%
\mathbf{z}=\mathbf{z}^{0}}=0  \tag{4.27}
\end{equation}%
are the Bethe ansatz equations for the Gaudin model. Using these equations
the eigenvalue, \ Eq.(4.17), for the Gaudin model now acquires the following
form:%
\begin{equation}
H_{i}(\mathbf{z}^{0})\mathcal{V(}\mathbf{x}^{0},\mathbf{z}^{0})=\frac{%
\partial }{\partial z_{i}}\ln \Phi _{k,n}(\mathbf{x}(\mathbf{z}),\text{ }%
\mathbf{z})\mid _{\mathbf{z}=\mathbf{z}^{0}}\mathcal{V(}\mathbf{x}^{0},%
\mathbf{z}^{0}).  \tag{4.28}
\end{equation}%
In the next subsection we shall sudy in some detail the Bethe ansatz \
Eq.s(4.28).This will allow us to obtain eigenvalues in \ Eq.(4.28)
explicitly.

\subsection{ Mathematics \ and physics of Bethe ansatz equations for XXX
Gaudin model according to Richardson. Connection with Veneziano model}

\bigskip Using \ Eq.(4.20) in (4.28) produces the following set of\ the
Bethe ansatz equations:%
\begin{equation}
\sum\limits_{i=1}^{n}\frac{m_{i}}{x_{j}-z_{i}}=\sum\limits_{\substack{ i=1 
\\ i\neq j}}^{k}\frac{2}{x_{j}-x_{i}},\text{ \ \ }j=1,...,k.  \tag{4.29}
\end{equation}%
To understand physical meaning of these equations \ we \ shall use
extensively results of\ two key papers by Richardson, Ref.s [68,69]. To
avoid duplications, and for the sake of space, our readers are encoraged to
read thoroughly these papers. \ Although \ originally they were written in
60ies having applications to nuclear physics in mind, they are no less
significant for condensed matter, Ref.[71], and atomic physcs, Ref.[73].
Because of this, only nuclear physics terminology will be occasionally used.
At the time of writing of these papers, QCD was still in its infancy.
Accordingly, no attempts were made to apply Richardson's results to QCD.
Recently, Ovchinnikov, Ref.[74], conjectured that the Richardson-Gaudin
equations can be useful for development of color superconductivity in QCD. A
comprehensive review of this topic is given in Ref.[75]. Incidentally, in
the same paper, Ref.[75], it is emphasized that such type of
superconductivity can exist \textsl{only} if the number of colors is \textbf{%
not} too large, e.g. N$_{c}$ =3. This fact is in accord with the remarks
made in Section 4.2. regarding the validity of WKB methods for K-Z equation
in the limit $N\rightarrow \infty .$These results clearly favour Abelian
reduction over ADS/CFT.

Below, we provide additional mathematically rigorous evidence supporting 't
Hooft's idea of Abelian reduction of QCD. These results should be considered
as complementary to that presented in Faddeev's paper, Ref.[27]. For this
purpose, following Richardson, Ref.[69], \ we consider the system of
interacting bosons described by the (pairing) Hamiltonian\footnote{%
In the paper with Sherman, Ref.[68], Richardson explains in detail how one
can map the fermionic (pairing) system into bosonic.}%
\begin{equation}
H=\sum\nolimits_{l}\varepsilon _{l}\hat{n}_{l}-\frac{g}{2}%
\sum\nolimits_{ll^{^{\prime }}}A_{l}^{+}A_{l}.  \tag{4.30}
\end{equation}%
Here we have $\hat{n}_{l}=\sum\limits_{\mathbf{k}(\varepsilon _{\mathbf{k}%
}=\varepsilon _{l})}a_{\mathbf{k}}^{+}a_{\mathbf{k}}$ , $A_{l}^{+}=\sum%
\limits_{\mathbf{k}(\varepsilon _{\mathbf{k}}=\varepsilon _{l})}a_{\mathbf{k}%
}^{+}a_{-\mathbf{k}}^{+}$ and $A_{l}=\sum\limits_{\mathbf{k}(\varepsilon _{%
\mathbf{k}}=\varepsilon _{l})}a_{-\mathbf{k}}a_{\mathbf{k}}.$ It is assumed
that the single-particle spectrum $\{\varepsilon _{l}\}$ is such that $%
\varepsilon _{l}<\varepsilon _{l+1}$ $\forall l$ and that the degeneracy of $%
l$-th level is $\Omega _{l}$ so that the sums (over $\mathbf{k}$ ) each
contain $\Omega _{l}$ terms. It is assumed furthermore that the system
possesses the time-reversal symmetry implying $\varepsilon _{\mathbf{k}%
}=\varepsilon _{-\mathbf{k}}$. The operators $a_{\mathbf{k}}^{+}$ and $a_{%
\mathbf{k}}$ obey usual commutation rules for bosons, i.e. $[a_{\mathbf{k}%
},a_{\mathbf{k}^{\prime }}^{+}]=\delta _{\mathbf{kk}^{\prime }}$. The sign
of the coupling constant, in principle, can be both positive and negative.
We shall work, however, with more physically interesting case of negative
coupling (so that $g$ in Eq.(4.30) is actually $\left\vert g\right\vert $).

An easy computation using commutation rule for bosons produces the following
results%
\begin{equation}
\lbrack \hat{n}_{l},A_{l^{^{\prime }}}^{+}]=2\delta _{ll^{\prime
}}^{{}}A_{l}^{+},  \tag{4.31a}
\end{equation}%
\begin{equation}
\lbrack A_{l},A_{l^{\prime }}^{+}]=2\delta _{ll^{\prime }}^{{}}(\Omega _{l}+2%
\hat{n}_{l}),  \tag{4.31b}
\end{equation}%
\begin{equation}
\lbrack \hat{n}_{l},A_{l^{^{\prime }}}^{{}}]=-2\delta _{ll^{\prime
}}^{{}}A_{l}^{{}}.  \tag{4.31c}
\end{equation}%
If we make a replacement \ of $\hat{n}_{l}$ in Eq.s(4.31a) and (4.31.c) by $%
\frac{\Omega _{l}}{2}+\hat{n}_{l}\equiv $ $\dfrac{\mathbf{\hat{n}}_{l}}{4}$
and keep the same notation in the r.h.s. of Eq.(4.31b) we shall arrive at
the $sl_{2}$ Lie algebra isomorphic to that given in Eq.(4.13). The same Lie
algebra was uncovered and used in our Part II for description of new models
describing Veneziano amplitudes. Because of this, we would like now to
demonstrate that the rest of arguments of Part II can be implemented in the
present context thus making the P-F model (which is derivative of the
Richardson-Gaudin XXX model) correct model related to Veneziano amplitudes.

Following Richardson, Ref.[69], we notice that the model described by the
Hamiltonian, Eq.(4.30), and algebra, Eq.s(4.31), admits two types of
excitations: those which are associated with the unpaired particles and
those with coupled pairs. The unpaired $\nu -$particle state is defined by
the following two equations%
\begin{equation}
\hat{n}\mid \varphi _{\nu }>=\nu \mid \varphi _{\nu }>,  \tag{4.32}
\end{equation}%
\begin{equation}
A_{l}\mid \varphi _{\nu }>=0\text{ }\forall l.  \tag{4.33}
\end{equation}%
Here, $\hat{n}=\sum\nolimits_{l}\hat{n}_{l}$ so that, in fact, 
\begin{equation}
\hat{n}_{l}\mid \varphi _{\nu }>=\nu _{l}\mid \varphi _{\nu }>  \tag{4.34}
\end{equation}%
and, therefore, $\nu =\sum\nolimits_{l}\nu _{l}$. Furthermore,%
\begin{equation}
H\mid \varphi _{\nu }>=\sum\nolimits_{l}\varepsilon _{l}\nu _{l}\mid \varphi
_{\nu }>.  \tag{4.35}
\end{equation}%
Following Richardson, we want to demonstrate that parameters $\varepsilon
_{l}$ in Eq.(4.35) can be identified with parameters $z_{l}$ in Bethe \ Eq.s
(4.29). Because of this, the eigenvalues for the P-F chain are obtained as
described \ in Section 4.2., that is 
\begin{equation}
E_{i}^{(\mathcal{P}-\mathcal{F})}=\frac{\partial }{\partial \varepsilon _{i}}%
\sum\nolimits_{l}\varepsilon _{l}\nu _{l}=\nu _{i}.  \tag{4.36}
\end{equation}%
These are the eigenvalues of $\hat{n}_{l}$ defined in Eq.(4.34).
Furthermore, this eigenvalue equation is \textsl{exactly the same} as was
used in Part II, Section 8, with purpose of reproducing Veneziano
amplitudes. Moreover, Eq.s(4.32) and (4.33) have the same mathematical
meaning as Eq.s (4.23) defining the Verma module. Because of this, we follow
Richardson's paper to describe this module in physical terms. By doing so
additional comparisons will be made between the results of Part II and works
of Richardson. Since the Hamiltonian, Eq.(4.30), describes two kinds of
particles: a) pairs of particles (whose total linear and angular momentum is
zero) and, b) unpaired particles (that is single particles which do not
interact with just described pairs), the total number of (quasi) particles
is $n=N+\nu \footnote{%
In Richardson's paper we find instead: $n=2N+\nu .$ This is, most likely, a
misprint as explained in the text.}$. Since we redefined the number operator
as $\frac{\Omega _{l}}{2}+\hat{n}_{l}\equiv $ $\dfrac{\mathbf{\hat{n}}_{l}}{4%
}\equiv \mathbf{\hat{N}}_{l},$ \ we expect that , once the correct state
vector describing excitations is found, Eq.(4.30) should be replaced by the
analogous equation for $\mathbf{\hat{N}}_{l}$ whose eigenvalues will be $%
\frac{\Omega _{l}}{2}+\nu _{l}$.\footnote{%
These amendments are not present in Richardson's paper but they are in
accord with its content.}

A simple minded way of creating such a state is by constructing the
following state vector $A_{l_{1}}^{+}\cdot \cdot \cdot A_{l_{N}}^{+}\mid
\varphi _{\nu }>$. This vector does not possess the needed symmetry of the
problem. To create the state vector (actually, the Bethe vector of the type
given by Eq.(4.24))\ of correct symmetry one should introduce a linear
combination \ of $A_{l}^{+}$ operators according to the following
prescription:%
\begin{equation}
B_{\alpha }^{+}=\sum\limits_{l}u_{\alpha }(l)A_{l}^{+},\text{ }\alpha
=1,...,N  \tag{4.37}
\end{equation}%
with constants $\ u_{\alpha }(l)$ to be determined below. The (unnormalized)
Bethe-type vectors are given then \ as $\mid \psi >=B_{1}^{+}\cdot \cdot
\cdot B_{N}^{+}\mid \varphi _{\nu }>$ and, accordingly, instead of
Eq.(4.35), we obtain 
\begin{equation}
H\mid \psi >=(\sum\nolimits_{l}\varepsilon _{l}\nu _{l})\mid \psi
>+[H,B_{1}^{+}\cdot \cdot \cdot B_{N}^{+}]\mid \varphi _{\nu }>.  \tag{4.38}
\end{equation}%
The task now lies in calculating the commutator and to determine the
constants $u_{\alpha }(l).$ Details can be found in Richardson's paper,
Ref.[69]. The final result looks \ as follows%
\begin{align}
H& \mid \psi >-E\mid \psi >  \tag{4.39} \\
& =\sum\limits_{\alpha =1}^{N}(\prod\limits_{\gamma \neq \alpha }B_{\gamma
}^{+})\sum\limits_{l}A_{l}^{+}[(2\varepsilon _{l}-E_{\alpha })u_{\alpha
}(l)+\sum\limits_{l^{\prime }}(\Omega _{l^{\prime }}+2\hat{n}_{l^{\prime
}})u_{\alpha }(l^{\prime })+4g\sum\limits_{\beta (\beta \neq \alpha
)}^{{}}M_{\beta \alpha }]\mid \varphi _{\nu }>.  \notag
\end{align}%
By requiring the r.h.s. of this equation to be zero, we arrive at the
eigenvalue equation 
\begin{equation}
H\mid \psi >=E\mid \psi >,\text{ where }E=\sum\limits_{l}\varepsilon _{l}\nu
_{l}+\sum\limits_{\alpha =1}^{N}E_{\alpha }.  \tag{4.40}
\end{equation}%
Furthermore, this requirement after several manipulations leads us to the
Bethe ansatz \ equations\footnote{%
It should be noted that in the original paper, Ref. [69], the sign in front
of the 3rd term in the l.h.s. is positive. This is because Richardson treats
both positive and negative couplings simultaneously. Equation (4.41a) is in
agreement with (3.24) of Richardson-Sherman paper, Ref.[68], where the case
of negative coupling (pairing) is treated.} 
\begin{equation}
\frac{1}{2g}+\sum\limits_{\beta (\beta \neq \alpha )}^{N}\frac{2}{E_{\beta
}-E_{\alpha }}-\sum\limits_{l=1}^{L}\frac{\Omega _{l}/2+\nu _{l}}{%
2\varepsilon _{l}-E_{\alpha }}=0,\text{ }\alpha =1,...,N,  \tag{4.41a}
\end{equation}%
as well to the explicit form of coefficients $u_{\alpha }(l):u_{\alpha
}(l)=1/(2\varepsilon _{l}-E_{\alpha })$ and that for the matrix elements $%
M_{\alpha ,\beta }$ (since, by construction, $u_{\alpha }(l)u_{\beta
}(l)=M_{\alpha ,\beta }u_{\alpha }(l)+M_{\beta ,\alpha }u_{\beta }(l)).$ In
the limit $g\rightarrow 0$ we expect $E_{\alpha }\rightarrow 2\varepsilon
_{l}$ $\ and$ $\Omega _{l}\rightarrow 0$ in accord with \ Eq.s(4.32)-(4.34).
Therefore, we conclude that $\frac{\Omega _{l}}{2}+\nu _{l}$ is an
eigenvalue of the operator \textbf{\^{N}}$_{l}$ acting on $\mid \psi >$ in
accord with remarks made before. In the opposite limit: $g\rightarrow \infty
,$ \ the system of Eq.s(4.41a) \ will coincide with Eq.(4.29) upon obvious
identifications: $x_{\alpha }\rightleftarrows E_{\alpha },2\varepsilon
_{l}\rightleftarrows z_{l},N\rightleftarrows k,L\rightleftarrows n$ and $%
\Omega _{l}/2+\nu _{l}\rightleftarrows m_{l}.$ Next, in view of Eq.s(4.36)
and (4.40) we obtain the following result for the occupation numbers:%
\begin{align}
\tilde{\Omega}_{i}& \equiv E_{i}^{(\mathcal{P-F})}=\frac{\partial }{\partial
\varepsilon _{i}}[\sum\limits_{l}\varepsilon _{l}\nu
_{l}+\sum\limits_{\alpha =1}^{N}E_{\alpha }]  \notag \\
& =\nu _{i}+\sum\limits_{\alpha =1}^{N}\frac{\partial E_{\alpha }}{\partial
\varepsilon _{i}}.  \tag{4.42}
\end{align}%
Based on the results just obtained, it should be clear that, actually, $%
E_{i}^{(\mathcal{P-F})}$ =$\nu _{i}+\frac{\Omega _{i}}{2}$ so that $\frac{%
\Omega _{i}}{2}=\sum\limits_{\alpha =1}^{N}\frac{\partial E_{\alpha }}{%
\partial \varepsilon _{i}}.$ Richardson, Ref.[69], \ cleverly demonstrated
that the combination $\sum\limits_{\alpha =1}^{N}\frac{\partial E_{\alpha }}{%
\partial \varepsilon _{i}}$ must be an integer.

Consider now a special case: $N=1$. Evidently, for this case, the derivative 
$\frac{\partial E_{\alpha}}{\partial\varepsilon_{i}}$ should \ also be an
integer. For different $\varepsilon_{i}^{\prime}s$ these may, in general, be
different integers. This fact has some physical significance to be explained
below.

To simplify matters, by analogy with theory of superconducting grains,
Ref.[71], we assume that $\ $the$\ $energy $\varepsilon_{i}$ can be written
as $\varepsilon_{i}=d(2i-L-1),$ $i=1,2,...,L.$ The adjustable parameter $d$
\ measures the level spacing for the unpaired \ particles in the limit $%
g\rightarrow0$. With such simplification, we obtain the following BCS-type
equation using \ Eq.s(4.41a) (for $N=1$):%
\begin{equation}
\sum\limits_{l=1}^{L}\frac{\tilde{\Omega}_{l}}{2\varepsilon_{l}-E}=\frac{1}{G%
},  \tag{4.43}
\end{equation}
where $G$ is the rescaled coupling constant. Such \ an equation was
discussed in the seminal paper by Cooper, Ref.[76], which paved a way to the
BCS theory of superconductivity. To solve this equation, let now $F(E)=\sum
\limits_{l=1}^{L}\tilde{\Omega}_{l}(2\varepsilon_{l}-E)^{-1}$ so that \
Eq.(4.43) is reduced to

\begin{equation}
F(E)=G^{-1}.  \tag{4.44}
\end{equation}
This equation can be solved graphically as depicted below, in Fig.1.

\begin{figure}[ptb]
\begin{center}
\includegraphics[width=2.50 in]{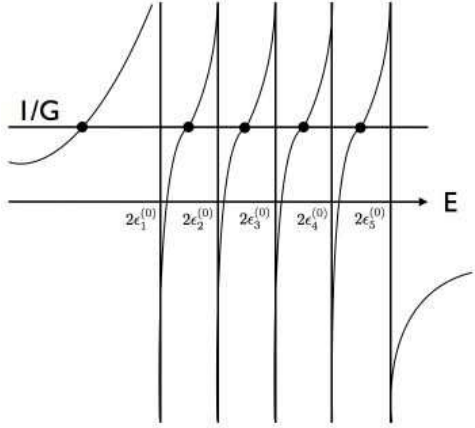}
\end{center}
\caption{Graphical solution of the Eq.(4.44)}
\end{figure}

As can be seen from Fig.1, solutions to this equation for $G=\infty $ \ can
be read off from the $x$ axis. In addition, if needed, \ for any $N\geq 1$
the system of Eq.s (4.41a) can be rewritten in a similar BCS-like form if we
introduce the renormalized coupling constant $G_{\alpha }$ via 
\begin{equation}
G_{\alpha }=G[1+2G\sum\limits_{\beta (\beta \neq \alpha )}^{N}\frac{1}{%
E_{\beta }-E_{\alpha }}]^{-1}\text{ }  \tag{4.45}
\end{equation}%
so that \ now we obtain:%
\begin{equation}
F(E_{\alpha })=G_{\alpha }^{-1},\alpha =1,...,N.  \tag{4.41b}
\end{equation}%
This system of equations can be solved iteratively, beginning with
Eq.(4.44). There is, however, better way of obtaing these solutions. In view
of Eq.s(4.19), (4.20) and (4.27) solutions $\{E_{\alpha }\}$ of \
Eq.(4.41.b) are the roots of the Lame$^{\prime }-$type function which is
obtained as solution of Eq.(4.19). Surprisingly, this fact known to
mathematicians for a long time has been recognized in nuclear physics
literature only very recently, e.g. read Ref.[77].

\subsection{Emergence of the Veneziano-like amplitudes \ as consistency
condition for $N=1$ solutions of K-Z equations. Recovery of pion-pion
scattering amplitude}

Since results for the Richardson-Gaudin (R-G) model are obtainable from the
corresponding solutions of K-Z equations, in this subsection we would like
to explain why $N=1$ solution of the Bethe-Richardon equations can be linked
with the Veneziano-like amplitudes describing the pion-pion scattering. In
doing so, we shall by pass the P-F model since, anyway, it is obtainable
from the R-G model.

Thus, we begin again with Eq.s (4.14),(4.15). We would like to look at
special class of solutions of Eq.(4.15) for which the parameter $\left\vert
J\right\vert $ in the Verma module, Eq.(4.23), is equal to one. This
corresponds exactly to the case $N=1$. Folloving Varchenko, Ref.[9], by
analogy with Eq.(4.20)\ we introduce function $\Phi (\mathbf{z},t)$ via 
\begin{equation}
\Phi (\mathbf{z},t)=\prod\limits_{1\leq i<j\leq L}(z_{i}-z_{j})^{\dfrac{%
m_{i}m_{j}}{\kappa }}\prod\limits_{l=1}^{L}(t-z_{l})^{-\dfrac{m_{l}}{\kappa }%
}.  \tag{4.46}
\end{equation}%
It is a multivalued function at \ the points of its singularities, i.e. at
the points $z_{1},...,z_{L}.$ Using this function, \ we define the set of
1-forms via%
\begin{equation}
\omega _{j}=\Phi (\mathbf{z},t)\frac{dt}{t-z_{j}},\text{ \ }j=1,...,L, 
\tag{4.47}
\end{equation}%
and the vector $\mathbf{I}^{(\gamma )}$ of integrals $\mathbf{I}^{(\gamma
)}=($I$_{1},...,$I$_{L})\equiv (\int\nolimits_{\gamma }\omega
_{1},...,\int\nolimits_{\gamma }\omega _{L})$ with $\gamma $ being a
particular Pochhammer countour: \ a double loop winding around \ any two
points $z_{\alpha }$, $z_{\beta }$ taken from the set $z_{1},...,z_{L}.$
Deatails can be found in Ref.s[9,63].

We want now to design the singular Verma module for the K-Z equations using
\ Eq.(4.23) and results just presented. Taking into account the following
known relations:%
\begin{equation*}
a)\text{ }ef^{k}v_{m}=k(m-k+1)f^{k-1}v_{m},\text{ and }b)\text{ }%
hf^{k}v_{m}=(m-2k)f^{k}v_{m}
\end{equation*}%
for the Lie algebra $sl_{2},$ also used in Part II, Section 8, and taking
into account that in the present ($N=1$) case the basis vectors $%
f^{J}v_{M}=f^{j_{1}}v_{m_{1}}\otimes \cdot \cdot \cdot \otimes
f^{j_{n}}v_{m_{n}}$ acquires the form: $f^{\mathbf{1}}v_{M}=v_{m_{1}}\otimes
\cdot \cdot \cdot \otimes fv_{m_{s}}\otimes \cdot \cdot \cdot \otimes
v_{m_{n}},s=1,...,L,$ provided that $m_{i}^{\prime }s$ are the same as in
Eq.(4.29) (or (4.46)), \ the singular vector for such a Verma module is
given by 
\begin{equation}
w(\gamma )=\sum\limits_{s=1}^{L}I_{s}v_{m_{1}}\otimes \cdot \cdot \cdot
\otimes fv_{m_{s}}\otimes \cdot \cdot \cdot \otimes v_{m_{n}}.  \tag{4.48}
\end{equation}%
In view of the Lie algebra relations just introduced, we obtain $e\cdot w=0$
or, explicitly, 
\begin{equation}
\sum\limits_{s=1}^{L}m_{s}I_{s}=0.  \tag{4.49}
\end{equation}%
Hence, for a fixed Pochhammer contour $\gamma $ there are $L-1$ independent
basis vectors $\{w^{i}\}$. They represent $L-1$ independent solutions of K-Z
equation of the type $k=1$ (or $N=1$). Let now $z_{i}^{\prime }s$ be ordered
in such a way that $z_{1}<\cdot \cdot \cdot <z_{L}.$ Furthermore, in view
their physical interpretation described in previous section, these $%
z_{i}^{\prime }s$ can be chosen to be equidistant. Consider then a special
set of Pochhamer contours $\{\gamma _{i}\}$ \ \ around points $z_{i}$ and $%
z_{i+1},$ $i=1,2,...,L-1,$ \ and consider the matrix $\mathbf{M}$ made of
integrals of the type \ $M_{j}^{i}=-\dfrac{m_{j}}{\kappa }%
\int\nolimits_{\gamma _{i}}\omega _{j}$ then, any ($k=1)-$ type solution $%
\phi ^{i}(i=1,2,...,L-1)$ of K-Z equation can be represented as 
\begin{equation}
\phi ^{i}=\sum\limits_{j}M_{j}^{i}w^{j},\text{ }i=1,2,...,L-1.  \tag{4.50}
\end{equation}%
From linear algebra it is known that in order for these K-Z solutions to be
independent \ we have to require that $\det \mathbf{M}$ $\neq 0.$ The proof
of this fact is given in the Appendix. Calculation of the determinant of $%
\mathbf{M}$ is described in detail in Ref.[9] so that we quote the result:%
\begin{equation}
\det \mathbf{M=\pm }\text{A}\frac{\Gamma (1-\frac{m_{1}}{\kappa })\cdot
\cdot \cdot \Gamma (1-\frac{m_{L}}{\kappa })}{\Gamma (1-\frac{\left\vert
M\right\vert }{\kappa })},  \tag{4.51}
\end{equation}%
where $\pm $A is some known constant\footnote{$\pm $A=$\prod\limits 
_{\substack{ 1\leq i,j\leq L  \\ (i\neq j)}}(z_{i}-z_{j})^{\dfrac{-m_{j}}{%
\kappa }}$} and $\Gamma (x)$ is Euler's gamma function. For $L=2$ without
loss of generality one can choose $z_{1}=0$ and $z_{2}=1,$ then \ in the
determinant thus obtained one easily can recognize the Veneziano-type $\pi
^{+}\pi ^{-}$ scattering amplitude used in the work by Lovelace, Ref.[78].
We discussed this amplitude previously in connection with mirror symmetry
issues in our work, Ref.[79]. This time, however, we would like to discuss
other topics.

In particular, we notice first that all mesons are made of two quarks.
Specifically, we have: $u\bar{d}$ for $\pi ^{+},d\bar{u}$ for $\pi ^{-}$ and 
$d\bar{d}$ for $\pi ^{0}.$ These are very much like the Cooper pairs with $q%
\bar{q}$ \ quark pairs contributing to the Bose condensate created as result
of spontaneous chiral symmetry breaking. As in the case of more familiar
Bose condensate, in addition to the ground state we expect to have a tower
of excited states made of such quark pairs. Experimentally, these are
interpreted as more massive mesons. Such excitations are ordered by their
energies, angular momentum and, perhaps, by other quantum numbers which can
be taken into account if needed. Color confinement postulate makes such a
tower infinite. Evidently, the Richardson-Gaudin model fits ideally this
qualitative picture. Eq.(4.44) describes excitations \ of such Cooper-like
pairs (even in the limit: $G\rightarrow \infty )$ as can be seen from Fig.1$%
. $In the P-F model the factor $\tilde{\Omega}_{i}$ plays effectively the
role of energy as \ already discussed in this work and in Part II.
Therefore, in view of \ Eq.(4.42), it is appropriate to write: $\tilde{\Omega%
}_{i}=f(E_{i}),$ with $E_{i}$ being the R-G energies. Although the explicit
form of such $f-$dependence may be difficult to obtain, for our purposes it
is sufficient only to know that such a dependence does exist. This then
allows us to make an identification: $\tilde{\Omega}_{i}\rightleftarrows 
\dfrac{m_{i}}{\kappa }$ consistent with Varchenko's results, e.g. compare
his Theorem 3.3.5 (page 35) with Theorem 6.3.2. (page 90) of Ref.[9]. But,
we already established that $\tilde{\Omega}_{i}$ is an integer, therefore, $%
\dfrac{m_{i}}{\kappa }$ should be also an integer. This creates some
apparent problems. For instance, when $\left\vert M\right\vert =\kappa $,
the determinant, $\det \mathbf{M,}$ becomes zero implying that solutions of
K-Z equation become interdependent. This fact has physical significance to
be discussed below and in Section 5. To do so we use some results from our
Part I. In particular, a comparison between 
\begin{equation}
\sin \pi z=\pi z\prod\limits_{k=1}^{\infty }(1-\left( \frac{k}{z}\right)
)(1+\left( \frac{k}{z}\right) )  \tag{4.52}
\end{equation}%
and 
\begin{equation}
\frac{1}{\Gamma (z)}=ze^{-Cz}\prod\limits_{k=1}^{\infty }(1+\left( \frac{k}{z%
}\right) )e^{-\dfrac{z}{k}},  \tag{4.53}
\end{equation}%
where $C$ is some known constant, \ tells us immediately that not only $%
\left\vert M\right\vert =\kappa $ will cause $\det \mathbf{M=}0$ but also $%
\left\vert M\right\vert =\kappa (k+1),k=0,1,2,...$ Accordingly, the
numerator in Eq.(4.51) will create poles whenever $\dfrac{m_{i}}{\kappa }=1.$
Existence of \ independent K-Z solutions is \textsl{not} destroyed if,
indeed, such poles do occur. These facts allow us to relabel $\dfrac{m_{i}}{%
\kappa }$ as $\alpha (s)$ (or $\alpha (t)$ or $\alpha (u)$, etc.) as it is
done in high energy physics with continuous parameters $s,t,u,...$replacing
discrete $i^{\prime }$s, different for different $\Gamma $ functions in the
numerator of \ Eq.(4.51). In the simplest case, this allows us to reduce the
determinant in Eq.(4.51) to the form used by Lovelace, i.e.%
\begin{equation}
\det \mathbf{M=-}\lambda \frac{\Gamma (1-\alpha (s))\Gamma (1-\alpha (t))}{%
\Gamma (1-\alpha (s)-\alpha (t))}.  \tag{4.54}
\end{equation}%
If, as usual, we parametrize $\alpha (s)=\alpha (0)+\alpha ^{\prime }s$,
then equation $1=\alpha (s)+\alpha (t)$ causes the $\det \mathbf{M}$ to
vanish. This also fixes the parameter $\alpha (0)$: $\alpha (0)=1/2.$ This
result was obtained \ by Adler long before sting theory emerged and is known
as Adler's \ selfconsistency condition, Ref.[80]. With such "gauge fixing",
one can fix the slope $\alpha ^{\prime }$\ as well \ if one notices that\
the experimental data allow us to make a choice: $1=\alpha (m_{\rho }^{2}).$%
This leads to: $\alpha ^{\prime }=\frac{1}{2m_{\rho }^{2}}\sim
0.885(Gev^{-2}),$ in accord with observations.

The obtained \ results are not limited to study of excitations of just one
"supeconducting" pair of quarks. In princile, any finite amount of such
pairs can be studied, e.g. see Ref.[81]. In such a case the result for $\det 
\mathbf{M}$ \ is expected to become considerably more complicated but
connections with one dimensional magnets still remain unchanged. We plan to
discuss these issues in future publications.

\section{ Random fragmentation and coagulation processes the
Poisson-Dirichlet distribution and Veneziano amplitudes}

\subsection{General facts about the Poisson-Dirichlet distribution}

In the Introduction, following Heisenberg, we posed a question: Is
combinatorics of observational data sufficient for recovery of underlying
unique microscopic model? That is, can we have complete understanding of
such a model based on information provided by combinatorics? As we
demonstrated, especially in Section 4 and in Ref.[37], this task is
impossible to accomplish without imposing additional constraints \ which,
normally, are not dictated by the combinatorics only. Even accounting for
such constraints, the obtained results could be in conflict with rigorous
mathematics and physical reality. Last but not the least, since Veneziano
amplitudes gave birth to string theory one can pose another question: Is
these Veneziano (or Veneziano-like) amplitudes, perhaps corrected to account
for particles with spin, contain enough information (analytical,
number-theoretic, combinatorial, etc.) allowing restoration of the
underlying microscopic model uniquely? \ \ In the most general case the
answer is: No!\ This happens in spite of the fact that \textsl{all}
amplitudes of high energy physics can be made out of linear combination of
Veneziano amplitudes (up to logarithmic corrections) as discussed in our
recent work, Ref.[4]. In the rest of this section we explain why this is so.

We begin with recalling some known auxiliary facts from the probability
theory. For instance, we recall that the stationary Maxwell distribution for
velocities of particles in the gas is of Gaussian-type. \ It can be obtained
\ as a stationary solution of the Boltzmann's dynamical equation maximizing
Boltzmann's entropy\footnote{%
As discussed in our work, Ref.[82], on the Poincar$e^{\prime }$ and
geometrization conjectures.}. The question arises: Is it possible to find
(discrete or continuous) dynamical equations which will provide known
probability distributions as stable stationary solutions? This task will
involve finding of dynamical equations along with the corresponding
Boltzmann-like entropies which \ will reach their maxima at respective
equilibria for these dynamical equations. We are certainly not in the
position in this work to discuss this problem in full generality. Instead,
we focus our attention only on processes described by the so called
Dirichlet distributions. These originate from the integral ( e.g. see
Eq.(2.8) of Part I) attributed to Dirichlet. It is given by 
\begin{equation}
\mathcal{D}(x_{1},...,x_{n+1})=\idotsint\limits_{\substack{ u_{1}\geq 0,...,%
\text{ }u_{n}\geq 0  \\ u_{1}+\cdot \cdot \cdot +u_{n}\leq 1}}%
u_{1}^{x_{1}-1}\cdot \cdot \cdot u_{n}^{x_{n}-1}(1-u_{1}-\cdot \cdot \cdot
-u_{n})^{x_{n+1}-1}du_{1}\cdot \cdot \cdot du_{n}.  \tag{5.1}
\end{equation}%
A random vector $(\mathbf{X}_{1},...,\mathbf{X}_{n})\in \mathbf{R}^{n}$ \
such that $\mathbf{X}_{i}\geq 0$ $\forall i$ and $\sum\limits_{i=1}^{n}$%
\textbf{\ \ }$\mathbf{X}_{i}$\textbf{\ }$\mathbf{=}1$\textbf{\ }is said to
be Dirichlet distributed with parameters ($x_{1},...,x_{n};x_{n+1}),$ e.g.
see Ref. [83], if the probability density function for $(\mathbf{X}_{1},...,%
\mathbf{X}_{n})$ is given by 
\begin{align}
P_{\mathbf{X}_{1},...,\mathbf{X}_{n}}(u_{1},...,u_{n})& =\frac{\Gamma
(x_{1}+\cdot \cdot \cdot +x_{n+1})}{\Gamma (x_{1})\cdot \cdot \cdot \Gamma
(x_{n+1})}u_{1}^{x_{1}-1}\cdot \cdot \cdot
u_{n}^{x_{n}-1}(1-\sum\limits_{i=1}^{n}u_{i})^{x_{n+1}-1}  \notag \\
& \equiv \frac{\Gamma (x_{1}+\cdot \cdot \cdot +x_{n+1})}{\Gamma
(x_{1})\cdot \cdot \cdot \Gamma (x_{n+1})}u_{1}^{x_{1}-1}\cdot \cdot \cdot
u_{n}^{x_{n}-1}u_{n+1}^{x_{n+1}-1},\text{ provided that \ }u_{n+1}  \notag \\
& =1-u_{1}-\cdot \cdot \cdot -u_{n}.  \tag{5.2}
\end{align}%
From these results it follows that Veneziano condition, Eq.(2.4), and
Veneziano amplitudes are inseparable from each other. Since Veneziano
condition is just restatement of energy-momentum conservation, such a
requirement should be applicable to whatever amplitude of high energy
physics. Not surprisingly, therefore, in Ref.[4], it is demonstrated that
this is indeed the case.

It is of interest to mention other uses of Dirichlet distributions beyond
that in high energy physics. For this purpose, to get a feeling of just
defined distribution, we notice the following peculiar aspects of this
distribution. For any discrete distribution, we know that the probability $%
p_{i}$ must be normalized, that is $\sum\nolimits_{i}p_{i}=1.$ Thus, the
Dirichlet distribution is dealing with averaging of the probabilities! Or,
better, is dealing with the problem of effectively selecting the most
optimal probability. The most primitive of these probabilities is the
binomial probability given by 
\begin{equation}
p_{m}=\left( 
\begin{array}{c}
n \\ 
m%
\end{array}%
\right) p^{m}(1-p)^{n-m},\text{ \ }m=0,1,2,....,n\text{.}  \tag{5.3}
\end{equation}%
If $X$ is random variable obeying this law of probability then, the
expectation \ $E(X)$ is calculated as 
\begin{equation}
E(X)=\sum\limits_{m=1}^{n}mp_{m}=np\equiv \mu .  \tag{5.4}
\end{equation}%
Consider such a distribution in the limit: $n\rightarrow \infty .$ In this
limit, if we write $p=\mu /n$ , then the Poisson distribution is obtained as 
\begin{equation}
p_{m}=\frac{\mu ^{m}}{m!}e^{-\mu }.  \tag{5.5}
\end{equation}%
Next, we notice that $m!=\Gamma (m+1),$ furthermore, we replace $m$ by the
real valued variable $\alpha $ and $\mu $ by $x$. This allows us to
introduce the gamma distribution \ with exponent $\alpha $ whose probability
density is 
\begin{equation}
p_{X}(x)=\frac{1}{\Gamma (\alpha )}x^{\alpha -1}e^{-x}  \tag{5.6}
\end{equation}%
for some gamma distributed random variable $X$. Finally, we would like to
demonstrate how the Dirichlet distribution can be represented through gamma
distributions. Since the gamma distribution originates from the Poisson
distribution, sometimes in literature the Dirichlet distribution is called
the Poisson-Dirichlet (P-D) distribution, Ref.[45]. To demonstrate
connection between the Dirichlet and gamma distributions is relatively easy.
Following Kingman, Ref.[45], consider a set of positive independent gamma
distributed random variables: $Y_{1},...,Y_{n+1}$ with exponents $\alpha
_{1},...,\alpha _{n+1}.$ Furthermore, consider \ $Y=Y_{1}+\cdot \cdot \cdot
+Y_{n+1}$ and construct \ a vector \textbf{u} with components: $u_{i}=\frac{%
Y_{i}}{Y}$. Then, since $\sum\nolimits_{i=1}^{n+1}u_{i}$ =$1,$ the
components of this vector are Dirichlet distributed and, in fact,
independent of $Y$. \ Details of the proof are based on results already
discussed in Part I and are given in Appendix.

Such described Dirichlet distribution is an equilibrium measure in various
fields ranging from spin glasses to computer science, from linguistics to
genetics, from forensic science to economics, etc. Many useful references\
involving these and other applications can be found in Ref.s[38-40].
Furtheremore, \ most of fragmentation and coagulation processes involve the
P-D distribution as their equilibrium measure. Some applications of general
theory of these processes to to nuclear and particle physics were initiated
in a series of papers by Mekjian, e.g. see Ref.[41] and references therein.
Alternative approach to the fragmentation-coagulation processes in high
energy physics was developed by Andersson, Ref.[42], and is known as the
Lund model. As results of our recent work, Ref.[4], indicate, the results of
Mekjian, Ref.[41], and that presented in this work are already fully
compatible with general theory of coagulation-fragmentation processes
discussed in Ref.s[38-40]. The interconnections between the Lund model and
general theory of coagulation-fragmentation processes remains to be
investigated. In the meantime, we would like to connect general results
presented in this subsection with those of Section 4. This is accomplished
below.

\subsection{ \ Quantum mechanics, hypergeometric\ functions and P-D
distribution}

In Ref.[2,3] we provided detailed explanation of the fact \ that all \
exactly solvable 2-body quantum mechanical problems involve different kinds
of special functions obtainable from the Gauss hypergeometric funcftion
whose integral representation is given by 
\begin{equation}
F(a,b,c;z)=\frac{\Gamma (c)}{\Gamma (b)\Gamma (c-b)}\int%
\limits_{0}^{1}t^{b-1}(1-t)^{c-b-1}(1-zt)^{-a}dt.  \tag{5.7}
\end{equation}%
As is well known, the disctete spectrum \ of all \ exactly solvable quantum
mechanical problems can be obtained only if one can find an appropriate set
of orthogonal polynomials related to this spectrum. Since all these
orthogonal polynomials are obtainable from Gauss hypergeometric function,
the question arises: Under what conditions on coefficients ($a,b$ and $c$)
can infinite hypergeometric series (whose integral representation is given
by Eq.(5.7)) be reduced to a finite (orthogonal) polynomial? This happens,
for instance, if we impose the \textsl{quantization condition}: $%
-a=0,1,2,... $ In such a case we can write $(1-zt)^{-a}=\sum%
\nolimits_{i=1}^{-a}(_{i}^{-a})(-1)^{i}(zt)^{i}$ \ and use this finite
expansion in Eq.(5.7). In view of Eq.(5.2), we obtain the convergent
generating function for the Dirichlet distribution. Hence, \textsl{all \
known quantum mechanical} \textsl{problems involving discrete spectrum} 
\textsl{are effectively examples of the P-D stochasic processes}.
Furthermore, from this point of view quantum mechanics ( also, quantum field
theory and string theory, e.g. see Ref.[4]) becomes just an applied theory
of the P-D stochastic processes. For hypergeometric functions of multiple
arguments this was demonstrated in Ref.[84] only quite recently.Other
arguments in favour of such an interpretation are developed in Ref.[4].

Next, we are\ still interested in the following. Given these observations,
can we include the determinantal formula, Eq.(4.51), into emerging
quantization scheme? \ Very fortunately, this can be done as explained in
the next subsection..

\subsection{\protect\bigskip Hypergeometric functions, Kummer series and
Veneziano amplitudes}

In view of just introduced new quantization condition, the question arises:
Is this the only condition reducing the hypergeometric function to a
polynomial ? More broadly: what conditions on coefficients $a,b$ and $c$ \
should be imposed so that the function $F(a,b,c;z)$ becomes a polynomial?
The answer to this question was provided by Kummer in the first half of 19th
century as discussed in Ref.[85]. Incidentally, in the case of K-Z equations
such a problem was solved only in 2007 in Ref.[86]. We would like to
summarize Kummer's results and to connect them with the determinantal
formula, Eq.(4.51).

According to general theory of hypergeometric equations \ of one variable\
discussed in Ref.[85], the infinite series for hypergeometric function
degenerates to a polynomial if one of the numbers%
\begin{equation}
a,b,c-a\text{ or \ }c-b  \tag{5.8}
\end{equation}
is an integer. This condition is equivalent to the condition that, at least
one of eight numbers $\pm(c-1)\pm(a-b)\pm(a+b-c),$ is an odd number.
According to general theory of hypergeometric functions of multiple
arguments discussed in Section 4, the $k=1$-type solutions can be obtained
using 1-forms given by Eq.(4.47) accounting for a singular module
constraint, Eq.(4.49), in the form given by Eq.(4.46). In the case of
Gauss-type hypergeometric functions, relations of the type given by
Eq.(4.49) were known already to Kummer. He found 24 interdependent
solutions. Evidently, this number is determined by the number of independent
Pochhamer contours as explained in Ref.s[9, 85]. Therefore, among these he
singled out 6 (generating these 24) and among these 6 he established that
every 3 of them are related to each other via equation of the type given by
Eq.(4.49).

We denote these 6 functions respectively as $u_{1},...,u_{6}.$ Then, we can
represent, say, $u_{2}$ and $u_{6}$ using $u_{1}$ and $u_{5}$ as the basis
set. We can do the same with $u_{1}$ and $u_{5}$ by representing them
through $u_{2}$ and $u_{6}$ and, finally, we can connect $u_{3}$ and $u_{4}$
with $u_{1}$ and $u_{5}.$ Hence, it is sufficient to consider, say, $u_{2}$
and $u_{6}.$ Thus, we obtain: 
\begin{equation}
\left( 
\begin{array}{c}
u_{2} \\ 
u_{6}%
\end{array}%
\right) =\left( 
\begin{array}{cc}
M_{1}^{1} & M_{2}^{1} \\ 
M_{1}^{2} & M_{2}^{2}%
\end{array}%
\right) \left( 
\begin{array}{c}
u_{1} \\ 
u_{5}%
\end{array}%
\right) ,  \tag{5.9}
\end{equation}%
with $M_{1}^{1}=\dfrac{\Gamma (a+b-c+1)\Gamma (1-c)}{\Gamma (a+1-c)\Gamma
(b-c+1)};$ $M_{2}^{1}=\dfrac{\Gamma (a+b+1-c)\Gamma (c-1)}{\Gamma (a)\Gamma
(b)};M_{1}^{2}=\dfrac{\Gamma (c+1-a-b)\Gamma (1-c)}{\Gamma (1-a)\Gamma (1-b)}%
;M_{2}^{2}=\dfrac{\Gamma (c+1-a-b)\Gamma (c-1)}{\Gamma (c-a)\Gamma (c-b)}.$
The determinant of this matrix becomes zero if either two rows or two
columns become the same. For instance, we obtain:%
\begin{equation}
\dfrac{\Gamma (a)\Gamma (b)}{\Gamma (c-1)}=\dfrac{\Gamma (a-c+1)\Gamma
(b-c+1)}{\Gamma (1-c)}\text{ and }\frac{\Gamma (c-a)\Gamma (c-b)}{\Gamma
(c-1)}=\frac{\Gamma (1-a)\Gamma (1-b)}{\Gamma (1-c)}.  \tag{5.10}
\end{equation}%
The condition $c=1$ \ in \ Eq.(5.10) causes two solutions of hypergeometric
equation to degenerate into one polynomial solution in accord with general
theory.

\medskip

\textbf{Appendix}

\ \ 

\ \textbf{A.Linear independence of solutions of K-Z equation}

\medskip

Linear independence of solutions of K-Z equation is based on the following
arguments. Consider change of the basis%
\begin{equation}
\mathbf{\tilde{e}}^{j}=A_{i}^{j}\mathbf{e}^{i}\text{ , }i,j=1,2,...,n 
\tag{A.1}
\end{equation}%
in $\mathbf{R}^{n}$. Using this result, consider the exterior product%
\begin{equation}
\mathbf{\tilde{e}}^{1}\wedge \cdot \cdot \cdot \wedge \mathbf{\tilde{e}}%
^{n}=[\det \mathbf{A]e}^{1}\wedge \cdot \cdot \cdot \wedge \mathbf{e}^{n}. 
\tag{A.2}
\end{equation}%
Next, suppose, that \ the vectors $\mathbf{\tilde{e}}^{j}$ are
lineraly-dependent. In particular, this means that 
\begin{equation}
\mathbf{\tilde{e}}^{n}=\alpha _{1}\mathbf{\tilde{e}}^{1}+\cdot \cdot \cdot
+\alpha _{n-1}\mathbf{\tilde{e}}^{n-1}  \tag{A.3}
\end{equation}%
for some nonzero $\alpha _{i}^{\prime }s.$\ Using this expansion in Eq.(A.2)
we obtain%
\begin{equation}
\mathbf{\tilde{e}}^{1}\wedge \cdot \cdot \cdot \wedge \mathbf{\tilde{e}}%
^{n-1}\wedge (\alpha _{1}\mathbf{\tilde{e}}^{1}+\cdot \cdot \cdot +\alpha
_{n-1}\mathbf{\tilde{e}}^{n-1})\equiv 0,  \tag{A.4}
\end{equation}%
implying $[\det \mathbf{A]=}0.$ Convesely, if $[\det \mathbf{A]\neq }0$%
\textbf{\ }then, vectors $\mathbf{\tilde{e}}^{j}$ \ are linearly independent.

\ \smallskip

\bigskip\textbf{B}. \textbf{Connections between the gamma and Dirichlet
distributions}

\smallskip

Using results of our Part I (especially Eq.(3.27)), such a connection can be
easily established. Indeed, consider $n+1$ independently \ distributed \
random gamma variables with exponents $\alpha_{1},...,\alpha_{n+1}$. The
joint probability density for such variables is given by 
\begin{equation}
p_{Y_{1}},..._{Y_{n+1}}(s_{1},...,s_{n+1})=\frac{1}{\Gamma(\alpha_{1})}%
\cdot\cdot\cdot\frac{1}{\Gamma(\alpha_{n+1})}s_{1}^{\alpha_{1}-1}\cdot
\cdot\cdot s_{n+1}^{\alpha_{n+1}-1}.  \tag{B.1}
\end{equation}
Let now $s_{i}=t_{i}t,$ \ where $t_{i}$ are chosen in such a way that $%
\sum\nolimits_{n=1}^{n+1}t_{i}=1.$ Then, using such a substitution in
Eq.(B.1), we obtain at once: 
\begin{equation}
p_{u_{1}},..._{u_{n+1}}(t_{1},...,t_{n+1})=[\int\limits_{0}^{\infty}t^{%
\alpha-1}e^{-t}]\frac{1}{\Gamma(\alpha_{1})}\cdot\cdot\cdot\frac{1}{%
\Gamma(\alpha_{n+1})}t_{1}^{\alpha_{1}-1}\cdot\cdot\cdot
t_{n+1}^{\alpha_{n+1}-1}  \tag{B.2}
\end{equation}
Since $\alpha=\alpha_{1}+\cdot\cdot\cdot+\alpha_{n+1}$, we also obtain: $%
\int\limits_{0}^{\infty}t^{\alpha-1}e^{-t}=\Gamma(\alpha_{1}+\cdot\cdot
\cdot+\alpha_{n+1}),$ implying that the density of probability given by
Eq.(B.2) is indeed of Dirichlet-type\ given by Eq.(5.2) of the main text.

\pagebreak

\bigskip\bigskip

\textbf{References}

\medskip

[1] \ \ P.Dirac, Lectures on Quantum Field Theory,

\ \ \ \ \ \ Yeshiva University Press, New York, 1996.

[2] \ \ A.Kholodenko, Heisenberg honeycombs solve Veneziano puzzle,

\ \ \ \ \ \ Int.Math.Forum 4 (2009) 441-509, hep-th/0608117.

[3] \ \ A.Kholodenko, Quantum signatures of Solar System dynamics,

\ \ \ \ \ \ arXiv: 0707.3992.

[4] \ \ A.Kholodenko, Landau's last paper and its impact on mathematics,

\ \ \ \ \ \ physics and other disciplines in new millenium, 

\ \ \ \ \ EJTP 5 (2008) 35-74, arXiv:0806.1064.

[5] \ \ A.Kholodenko, New strings for old Veneziano amplitudes I\textit{.}

\ \ \ \ \ \ Analytical treatment, J.Geom.Phys.55 (2005) 50-74.

[6] \ \ A.Kholodenko, New strings for old Veneziano amplitudes II\textit{.}

\ \ \ \ \ \ Group-theoretic treatment\textit{, }J.Geom.Phys.56
(2006)1387-1432.

[7] \ \ A.Kholodenko, New strings for old Veneziano amplitudes III\textit{.}

\ \ \ \ \ \ \ Symplectic treatment\textit{,} J.Geom.Phys.56 (2006) 1433-1472.

[8] \ \ \ N.Reshetikhin and A.Varchenko, Quasiclassical asymptotics of%
\textit{\ }

\ \ \ \ \ \ \ solutions of KZ equations, in Geometry, Topology and Physics
for

\ \ \ \ \ \ \ Raoul Bott, pp. 293-322, International Press, Boston, 1995.

[9] \ \ \ A.Varchenko, Special functions, KZ type equations, and\textit{\ }

\ \ \ \ \ \ \ \ representation theory, AMS Publishers, Providence, RI, 2003.

[10] \ \ P.Collins, Introduction to Regge Theory and High Energy Physics,

\ \ \ \ \ \ \ \ Cambrodge University Press, Cambridge, 1977.

[11] \ \ D. Bardin and G. Passarino, The Standard Model in the Making,

\ \ \ \ \ \ \ \ Clarendon Press, Oxford, 1999.

[12] \ \ R. Leigh, D. Minic, A.Yelnikov, Solving pure Yang-Mills in 2+1

\ \ \ \ \ \ \ \ dimensions, PRL 96 (2006) 222001.

[13] \ \ L. Freidel, R. Leigh, D. Minic, A. Yelnikov, On the spectrum of

\ \ \ \ \ \ \ \ pure Yang-Mills theory, arXiv:0801.1113.

[14] \ \ J. Maldacena, The Large N Limit of Superconformal Field Theories

\ \ \ \ \ \ \ \ and Supergravity, Adv.Theor.Math.Phys. 2 (1998) 231-252.

[15] \ \ M. Benna, I. Klebanov, Gauge-string dualities and some applications,

\ \ \ \ \ \ \ \ arXiv: 0803.1315.

[16] \ \ \ S. Brodsky, ADS/CFT and QCD, hep-th/0702205.

[17] \ \ A.Kholodenko, Boundary CFT, limit sets of Kleinian groups and

\ \ \ \ \ \ \ \ holography, J.Geom.Phys.35 (2000) 193-238.

[18] \ \ G. 't Hooft, Topology of the gauge condition and new confinement

\ \ \ \ \ \ \ \ phases in non-Abelian gauge theories,

\ \ \ \ \ \ \ \ Nucl.Phys. B190 (1981) 455-478.

[19] \ \ Y. Nambu, Strings, monopoles and gauge fields,

\ \ \ \ \ \ \ \ Phys.Rev.D 10 (1974) 4262-4268.

[20] \ \ T. Suzuki, I. Yotsuyanagi, \ Possible evedence for Abelian dominance

\ \ \ \ \ \ \ \ in quark confinement, Phys.Rev. D 42 (1990) 4257-4260.

[21] \ \ J. Stack, S. Neiman, R.Wensley, String Tension from Monopoles

\ \ \ \ \ \ \ \ in SU(2) Lattice Gauge Theory, Phys Rev.D 50 (1994)
3399-3405.

[22] \ \ Y. Cho, Restricted gauge theory, Phys. Rev.D 21 (1980) 1080-1088.

[23] \ \ Y. Cho, D. Pak, Monopole condensation in SU(2) QCD,

\ \ \ \ \ \ \ \ Phys.Rev.D 65 (2002) 074027.

[24] \ \ K-I. Kondo, Gauge-invariant gluon mass, infrared Abelian dominance,

\ \ \ \ \ \ \ \ and stability of magnetic vacuum, Phys. Rev. D 74 (2006)
125003.

[25] \ \ K-I. Kondo, A. Ono, A. Shibata, T. Shinohara, T. Murakami,

\ \ \ \ \ \ \ \ Glueball mass from quantized knot solitons and

\ \ \ \ \ \ \ \ gauge-invariant gluon mass, J.Phys. A 39 (2006) 13767--13782.

[26]\ \ \ K-I. Kondo, Magnetic monopoles and center vortices

\ \ \ \ \ \ \ \ as gauge-invariant topological defects simultaneously
responsible

\ \ \ \ \ \ \ \ for confinement, arXiv: 0802.3829.

[27] \ \ L. Faddeev, Knots as possible excitations of the quantum

\ \ \ \ \ \ \ \ Yang-Mills fields, arXiv: 0805.1624.

[28] \ \ D. Auckley, L. Kapitanski, J. Speight, Geometry and analysis

\ \ \ \ \ \ \ \ in nonlinear sigma models, St.Petersburg Math. J. 18 (2007)
1-19.

[29] \ \ \ Y.Cho, D.Pak, P.Zhang, New interpretation of Skyrme theory,

\ \ \ \ \ \ \ \ \ arXiv: hep-th/0404181

[30] \ \ \ D. Auckly, L. Kapitanski, Analysis of S$^{2}$-valued maps and

\ \ \ \ \ \ \ \ \ Faddeev's model, Comm.Math.Phys.256 (2005) 611-629.

[31] \ \ \ M. Asorey, F. Falceto, H. Sierra, Chern-Simons theory and

\ \ \ \ \ \ \ \ \ BCS superconductivity, Nucl.Phys.B 622 (2002) 593-614.

[32] \ \ \ N. Dorey, A spin chain from string theory, arXiv: 0805.4387.

[33] \ \ \ S. Gubser, I. Klebanov, A.Polyakov, Gauge theory correlators

\ \ \ \ \ \ \ \ \ from non-critical string theory, Phys.Lett.B 428 (1998)
105-114.

[34] \ \ \ J. Minahan, K. Zarembo, The Bethe ansatz for $\mathcal{N}$=4 super

\ \ \ \ \ \ \ \ \ Yang-Mills, JHEP 0303 (2003) 013.

[35] \ \ \ M. Kruczenski, Spin chains and string theory,

\ \ \ \ \ \ \ \ \ PRL 93 (2004) 161602.

[36] \ \ \ \ A. Cootrone, L.Martucci, J.Pons, P.Talavera, Heavy hadron

\ \ \ \ \ \ \ \ \ \ spectra from spin chains and strings, JHEP 05 (2007) 027.

[37] \ \ \ \ A. Kholodenko, New strings for old Veneziano amplitudes IV.

\ \ \ \ \ \ \ \ \ \ Connections with spin chains and other stochastic
systems,

\ \ \ \ \ \ \ \ \ \ arxiv: 0805.0113.

[38] \ \ \ \ J. Bertoin, Random fragmentation and coagulation processes,

\ \ \ \ \ \ \ \ \ \ Cambridge University Press, Cambridge U.K., 2006.

[39] \ \ \ \ J. Pitman, Combinatorial stochasic processes,

\ \ \ \ \ \ \ \ \ \ Springer-Verlag, Berlin, 2006.

[40] \ \ \ \ R. Arratia, A.Barbour and S.Tavare, Logarithmic Combinatorial%
\textit{\ }

\textit{\ \ \ \ \ \ \ \ \ \ }Structures: A Probabilistic Approach\textit{, }

\ \ \ \ \ \ \ \ \ \ European Mathematical Society, Z\"{u}rich, 2003.

[41] \ \ \ \ A. Mekjian, Model for studing brancing processes, multiplicity

\ \ \ \ \ \ \ \ \ \ \ distribution and non-Poissonian fluctuations in heavy
-ion collisions\textit{,}

\ \ \ \ \ \ \ \ \ \ PRL 86 (2001) 220-224.

[42] \ \ \ \ B. Andersson, The Lund Model, Cambridge University Press,

\ \ \ \ \ \ \ \ \ \ Cambridge, UK, 1998.

[43] \ \ \ \ R. Stanley, Combinatorics and Commutative Algebra, Birkh\"{a}%
user,

\ \ \ \ \ \ \ \ \ \ Boston, 1996.

[44] \ \ \ \ \ S. Ghorpade and G.Lachaud, Hyperplane sections of Grassmannian

\ \ \ \ \ \ \ \ \ \ \ and the number of MDS linear codes, Finite Fields \&

\ \ \ \ \ \ \ \ \ \ \ Their Applications 7\textbf{\ (}2001) 468-476\textbf{.}

[45] \ \ \ \ \ J. Kingman, Poisson processes, Clarendon Press, Oxford, 1993.

[46] \ \ \ \ \ R.Stanley, Enumerative combinatorics, Vol.1,

\ \ \ \ \ \ \ \ \ \ \ Cambridge University Press, Cambridge, 1999.

[47] \ \ \ \ \ S.Mohanty, Lattice path counting and applications,

\ \ \ \ \ \ \ \ \ \ \ \ Academic Press, New York, 1979.

[48] \ \ \ \ \ A.Vershik, Asymtotic Combinatorics With Applications to

\ \ \ \ \ \ \ \ \ \ \ Mathematical Physics, LNM 1815, Springer-Verlag,
Berlin, 2003.

[49] \ \ \ \ \ F.Goodman, P.de la Harpe, V.Jones, Coxeter Graphs and Towers

\ \ \ \ \ \ \ \ \ \ \ of Algebras, Springer-Verlag, Berlin, 1989.

[50] \ \ \ \ \ R.Bott and L.Tu, Differential forms in algebraic topology,

\ \ \ \ \ \ \ \ \ \ \ Springer-Verlag, Berlin, 1982.

[51] \ \ \ \ \ J.Schwartz, Differential Geometry and Topology, Gordon

\ \ \ \ \ \ \ \ \ \ \ and Breach, Inc., New York, 1968.

[52] \ \ \ \ \ M.Stone, Supersymmetry and quantum mechanics of spin,

\ \ \ \ \ \ \ \ \ \ \ Nucl.Phys. B 314 (1989) 557-586.

[53] \ \ \ \ \ O. Alvarez, I \ Singer and P.Windey, Quantum mechanics and

\ \ \ \ \ \ \ \ \ \ \ the\ geometry of the Weyl character formula\textit{,}

\ \ \ \ \ \ \ \ \ \ \ Nucl.Phys.B 337 (1990) 467-486.

[54] \ \ \ \ \ A. Polyakov, Gauge Fields and Strings,

\ \ \ \ \ \ \ \ \ \ \ Harwood Academic Publ., New York, 1987.

[55] \ \ \ \ \ A. Polychronakos, Exact spectrum of SU(n) spin chain with

\ \ \ \ \ \ \ \ \ \ \ inverse square exchange\textit{, }Nucl.Phys. B 419
(1994) 553-566.

[56] \ \ \ \ \ \ H. Frahm, Spectrum of a spin chain with invese square
exchange,

\ \ \ \ \ \ \ \ \ \ \ J.Phys.A 26 (1993) L 473-479.

[57] \ \ \ \ \ A. Polychronakos, Generalized statistics in one dimension,

\ \ \ \ \ \ \ \ \ \ \ hep-th/9902157.

[58] \ \ \ \ \ A. Polychronakos, Physics and mathematics of Calogero
particles,

\ \ \ \ \ \ \ \ \ \ \ hep-th/0607033.

[59] \ \ \ \ \ K. Hikami, Yangian symmetry and Virasoro character in a
lattice

\ \ \ \ \ \ \textit{\ \ \ \ \ }spin \ system with long -range interactions,

\ \ \ \ \ \ \ \ \ \ \ Nucl.Phys.B 441 (1995) 530-548.

[60] \ \ \ \ \ E. Melzer, The many faces of a character\textit{,}
hep-th/9312043.

[61] \ \ \ \ \ R. Kedem, B. McCoy and E. Melzer, The sums of Rogers,

\ \ \ \ \ \ \ \ \ \ \ Schur and\textit{\ \ }Ramanujian and the Bose-Fermi
correspondence in 1+1

\ \ \ \ \ \ \ \ \ \ \ dimensional quantum field theory, hep-th/9304056%
\textit{.}

[62] \ \ \ \ \ A.Tsvelik, Quantum Field Theory in Condensed Matter Physics,

\ \ \ \ \ \ \ \ \ \ \ Cambridge University Press, Cambridge U.K., 2003.

[63] \ \ \ \ \ \ \ \ P. Etingof, I. Frenkel and A. Kirillov Jr., Lectures on
Representation\textit{\ }

\ \ \ \ \ \ \ \ \ \ \ Theory and Knizhnik-Zamolodchikov Equations,

\ \ \ \ \ \ \ \ \ \ \ AMS Publishers, Providence, R.I., 1998.

[64] \ \ \ \ \ \ A.Chervov and D.Talalaev, Quantum spectral curves, quantum

\ \ \ \ \ \ \ \ \ \ \ \ integrable systems and the geometric Langlands
correspondence,

\ \ \ \ \ \ \ \ \ \ \ \ arXiv: hep-th/0604128.

[65] \ \ \ \ \ \ \ E.Frenkel, Langlands Correspondence For Loop Groups,

\ \ \ \ \ \ \ \ \ \ \ \ \ Cambridge University Press, Cambridge, U.K., 2007.

[66] \ \ \ \ \ \ \ E.Frenkel and E.Witten, Geometric endoscopy and mirror

\ \ \ \ \ \ \ \ \ \ \ \ \ symmetry, arXiv: 0710.5939.

[67] \ \ \ \ \ \ \ M.Gaudin, La function D'onde de Bethe\textit{,}

\ \ \ \ \ \ \ \ \ \ \ \ \ Masson, Paris, 1983.

[68] \ \ \ \ \ \ \ R.Richardson and N.Sherman, Exact eigenvalues of the

\ \ \ \ \ \ \ \ \ \ \ \ \ pairing-force\textit{\ }\ Hamiltonian,
Nucl.Phys.52 (1964) 221-243.

[69] \ \ \ \ \ \ \ R.Richardson, Exactly solvable many-boson model,

\ \ \ \ \ \ \ \ \ \ \ \ JMP 9 (1968) 1327-1348.

[70] \ \ \ \ \ \ \ N.Vilenkin, Special functions and theory of group
representations,

\ \ \ \ \ \ \ \ \ \ \ \ \ Nauka, Moscow, 1991.

[71] \ \ \ \ \ \ \ J.Dukelsky, S.Pittel and G.Sierra, Exactly solvable%
\textit{\ }

\ \ \ \ \ \ \ \ \ \ \ \ \ Richardson-Gaudin \ models for many-body\textit{\ }

\ \ \ \ \ \ \ \ \ \ \ \ \ quantum systems, Rev.Mod.Phys.76 (2004) 643-662.

[72] \ \ \ \ \ \ \ A.Balantekin, T.Dereli and Y.Pehlivan, Exactly solvable
pairing

\ \ \ \ \ \ \ \ \ \ \ \ \ model using an extension of Richardson-Gaudin
approach,

\ \ \ \ \ \ \ \ \ \ \ \ \ Int.J.Mod.Phys.E 14 (2005) 47-55.

[73] \ \ \ \ \ \ \ A.Ushveridze, Quasi-exactly solvable models in quantum

\ \ \ \ \ \ \ \ \ \ \ \ \ mechanics, IOP Publishing Ltd., Philadelphia, 1994.

[74] \ \ \ \ \ \ \ A.Ovchinnikov, Exactly solvable discrete BCS-type
Hamiltonians

\ \ \ \ \ \ \ \ \ \ \ \ \ and the six-vertex model, Nucl.Phys. B707 (2002)
362-378.

[75] \ \ \ \ \ \ \ M. Alford, A. Schitt, K. Rajagopal and Th.Schafer,

\ \ \ \ \ \ \ \ \ \ \ \ \ Color superconductivity in quark matte\textit{r},

\ \ \ \ \ \ \ \ \ \ \ \ \ arXiv: 0709.4635.

[76] \ \ \ \ \ \ \ L.Cooper, Bound electron pairs in a degenerate electron
gas,

\ \ \ \ \ \ \ \ \ \ \ \ \ Phys.Rev.104 (1956) 1189-1180.

[77] \ \ \ \ \ \ \ A.Balantekin, J.de Jesus and Y.Pehlivan, Spectra and

\ \ \ \ \ \ \ \ \ \ \ \ \ symmetry in \textit{\ }nuclear pairing, Phys.Rev.
C 75 (2007) 064304.

[78] \ \ \ \ \ \ \ C. Lovelace, A novel application of Regge trajectories,

\ \ \ \ \ \ \ \ \ \ \ \ \ Phys.Lett.B 28 (1968) 264-267.

[79] \ \ \ \ \ \ \ A.Kholodenko, Traces of mirror symmetry in Nature,

\ \ \ \ \ \ \ \ \ \ \ \ \ International Math.Forum 3 (2008) 151-184.

[80] \ \ \ \ \ \ \ \ S. Adler, Consistency conditions on a strong
interactions

\ \ \ \ \ \ \ \ \ \ \ \ \ implied by a partially conserved axial-vector
current,

\ \ \ \ \ \ \ \ \ \ \ \ \ Phys.Rev. 137 (1965)1022-1027.

[81] \ \ \ \ \ \ \ K. Iida, G. Baym, Superfluid phases in quark model:

\ \ \ \ \ \ \ \ \ \ \ \ \ Ginzburg-Landau theory and color neutrality,

\ \ \ \ \ \ \ \ \ \ \ \ \ Phys.Rev.D 63 (2001) 074018.

[82] \ \ \ \ \ \ \ \ A. Kholodenko, Towards physically motivated proofs

\ \ \ \ \ \ \ \ \ \ \ \ \ of the Poincare and geometrization conjectures,

\ \ \ \ \ \ \ \ \ \ \ \ \ J.Geom.Phys.58 (2008) 259-290.

[83] \ \ \ \ \ \ \ \ N.Balakrisnan, V.Nevzorov, A Primer on Statistical

\ \ \ \ \ \ \ \ \ \ \ \ \ \ Distributions, Wiley Interscience Inc., NY, 2003.

[84] \ \ \ \ \ \ \ \ A. Lijoi, E.Regazzini, Means of a Dirichlet process

\ \ \ \ \ \ \ \ \ \ \ \ \ and multiple hypergeometric functions,

\ \ \ \ \ \ \ \ \ \ \ \ Ann.Probability 32 (2004) 1469-1495.

[85] \ \ \ \ \ \ H.Bateman and A.Erdelyi, Higher Transcendental Functions,

\ \ \ \ \ \ \ \ \ \ \ \ Vol.1. McGraw Hill, NY, 1953.

[86] \ \ \ \ \ \ G.Felder and A.Veselov, Polynomial solutions

\ \ \ \ \ \ \ \ \ \ \ \ of the Knizhnik-Zamolodchikov equations and

\ \ \ \ \ \ \ \ \ \ \ \ Schur-Weyl duality, Intern.Math.Res. Notices

\ \ \ \ \ \ \ \ \ \ \ \ Article ID rnm046, 2007.

\bigskip

\bigskip

\bigskip

\end{document}